\documentclass[prl,
               twocolumn,
               final,
               reprint,
               amsmath,
               amssymb,
               superscriptaddress,
               showpacs,
               10pt,
               hidelinks,
               numbers,
               sort&compress,
               floats,
               nobibnotes,
               balancelastpage,
               longbibliography]{revtex4-2}
\usepackage{subfiles}               
\usepackage[journal=sciadvs,
            euler=false,  
            nag=true,     
            todo=false,
            ]{my_paper_2}

\usepackage{graphicx, color} 

\usepackage[normalem]{ulem}          

\usepackage{hyperref}
\hypersetup{colorlinks=true,
            pdfpagemode=None,
            linkcolor=black,
            citecolor=blue,
            urlcolor=blue}

\usepackage[utf8]{inputenc}

\usepackage[caption=false]{subfig}
\usepackage{amssymb}
\usepackage{amsmath}
\usepackage{commath}
\usepackage{graphicx,bm}
\usepackage{verbatim}

\newcommand{\myurl}[1]{\href{https://#1}{\textsf{#1}}}

\newcommand{\noacronym}[2]{\newacronym{#1}{#2}{<undefined>}\glsunset{#1}}
\newacronym{NSF}{\mysc{nsf}}{National Science Foundation}
\newacronym{HPC}{\mysc{hpc}}{high-performance computing}
\newacronym{GPU}{\mysc{gpu}}{graphics processing unit}
\newacronym{CPU}{\mysc{cpu}}{central processing unit}
\glsunset{CPU}
\newacronym{DFT}{\mysc{dft}}{density functional theory}
\newacronym{TDDFT}{\mysc{tddft}}{time-dependent density functional theory}
\newacronym{FLOPS}{\mysc{flops}}{floating point operations per second}
\noacronym{LUMI}{\mysc{lumi}}
\noacronym{RIKEN}{\mysc{riken}}
\noacronym{EuroHPC}{E\textup{uro} \mysc{hpc}}
\noacronym{AMD}{\mysc{amd}}
\newacronym{GCD}{\mysc{gcd}}{graphics compute die}
\newacronym{ELPA}{\mysc{elpa}}{Eigenvalue soLvers for Petaflop Applications}
\noacronym{CSC}{\mysc{csc}}
\newacronym{HIP}{\mysc{hip}}{Heterogeneous-Compute Interface for Portability}
\newacronym{HBM2}{\mysc{hbm2}}{high-bandwidth memory}
\newacronym{PDE}{\mysc{pde}}{partial differential equation}
\newacronym{ABM}{\mysc{abm}}{Adams-Bashforth-Moulton}
\newacronym{LDA}{\mysc{lda}}{local density approximation}
\newacronym{SLDA}{\mysc{slda}}{superfluid local density approximation}
\newacronym{TDSLDA}{\mysc{tdslda}}{time-dependent superfluid local density approximation}
\newacronym{QMC}{\mysc{qmc}}{quantum Monte Carlo}
\newacronym{qpwf}{\textup{qpwf}}{quasiparticle wavefunction}
\newacronym{GPE}{\mysc{gpe}}{Gross-Pitaevskii equation}
\newacronym{BEC}{\mysc{bec}}{Bose-Einstein condensate}
\newacronym{BCS}{\mysc{bcs}}{Bardeen Cooper Schrieffer}
\newacronym{MPI}{\mysc{mpi}}{message-passing interface}
\newacronym{HFB}{\mysc{hfb}}{Hartee-Fock-Bogoliubov}
\newacronym{BdG}{\mysc{b}\textup{d}\mysc{g}}{Bogoliubov-de Gennes}
\newacronym{FFT}{fft}{fast Fourier transform}
\newacronym{DVR}{\mysc{dvr}}{discrete variable representation}
\noacronym{hipFFT}{\textup{hip}\mysc{fft}}
\newacronym{W-SLDA}{\mysc{w-slda}}{Warsaw \gls{SLDA}}
\newacronym{API}{\mysc{api}}{application programming interface}
\glsunset{W-SLDA}
\noacronym{NVIDIA}{\mysc{nvidia}}
\noacronym{MI250}{\mysc{mi}250}
\noacronym{MI250X}{\mysc{mi}250\mysc{x}}
\noacronym{CDNA}{\mysc{cdna}}
\noacronym{ROCm}{\mysc{roc}\textup{m}}

\noacronym{A100}{\mysc{a100}}
\noacronym{RAVEN}{\mysc{raven}}
\newacronym{MPCDF}{\mysc{mpcdf}}{Max Planck Computing and Data Facility}
\newacronym{UFG}{\mysc{ufg}}{unitary Fermi gas}
\newacronym{EVP}{\mysc{evp}}{eigenvalue problem}
\noacronym{ScaLAPACK}{\mysc{s}\textup{ca}\mysc{lapack}}

\newcommand{\term}[1]{\textit{#1}}

\renewcommand{\vF}{v_{F}}

\newcommand{\kF}{k_{F}}
\newcommand{\kFa}{\kF a}
\newcommand{\eF}{\varepsilon_{F}}

\newcommand{\Eflow}{E_{\textrm{flow}}}
\newcommand{\Econd}{E_{\textrm{cond}}}
\newcommand{\lvor}{l_{\text{vor}}}
\newcommand{\Lvor}{L_{\text{vor}}}
\renewcommand{\lvor}{l}
\renewcommand{\Lvor}{L}
\renewcommand{\vect}[1]{\boldsymbol{#1}}
\let\Im\relax
\DeclareMathOperator{\Im}{Im}

\usepackage[caption=false]{subfig}
\newcommand{\labelphantom}[1]{\subfloat{\label{#1}}}

\begin{document}

\title[Fermionic Turbulence with HPC]
      {Fermionic Quantum Turbulence:
       Pushing the Limits of High-Performance Computing}

\author{Gabriel Wlaz\l{}owski}\email{gabriel.wlazlowski@pw.edu.pl}
\affiliation{Faculty of Physics, Warsaw University of Technology, Ulica Koszykowa 75, 00-662 Warsaw, Poland}
\affiliation{Department of Physics, University of Washington, Seattle, Washington 98195--1560, USA}

\author{Michael McNeil Forbes}\email{m.forbes@wsu.edu}
\affiliation{Department of Physics and Astronomy, Washington State University, Pullman, WA 99164, USA}
\affiliation{Department of Physics, University of Washington, Seattle, Washington 98195--1560, USA}

\author{Saptarshi Rajan Sarkar}\email{saptarshi.sarkar@wsu.edu}
\affiliation{Department of Physics and Astronomy, Washington State University, Pullman, WA 99164, USA}

\author{Andreas Marek}\email{andreas.marek@mpcdf.mpg.de}
\affiliation{\Gls{MPCDF}, 85741 Garching near Munich, Germany}

\author{Maciej Szpindler}\email{m.szpindler@cyfronet.pl}
\affiliation{Academic Computer Centre CYFRONET, University of Science and Technology in Cracow, Nawojki 11, 30-950 Cracow, Poland}

\date{\today}

\begin{abstract}
  Ultracold atoms provide a platform for analog quantum computer capable of simulating the
  quantum turbulence that underlies puzzling phenomena like pulsar glitches in rapidly spinning neutron stars.
  Unlike other platforms like liquid helium, ultracold atoms have a viable theoretical framework for dynamics, but simulations push the edge of current classical computers.
  We present the largest simulations of fermionic quantum turbulence to date and explain the computing technology needed, especially improvements in the \gls{ELPA} library that enable us to diagonalize matrices of record size (millions by millions).
  We quantify how dissipation and thermalization proceed in fermionic quantum
  turbulence by using the internal structure of vortices as a new probe of the local effective temperature.
  All simulation data and source codes are made available to facilitate rapid scientific progress in the field of ultracold Fermi gases.
\end{abstract}
\glsreset{ELPA}

\maketitle

\section{Significance}
Accurate simulations of quantum systems are challenging for computational physics, yet essential for developing new technologies.
    The convergence of theory, algorithms, and supercomputers allows us to diagonalize million-by-million matrices and evolve millions of coupled partial differential equations to simulate complex phenomena in systems consisting of tens of thousands of superfluid fermions.
    Our results demonstrate key aspects of fermionic quantum turbulence that simpler models do not capture.
    We provide both data and open-source codes, establishing a benchmark for ultracold atom experiments to validate the theory.
    This framework will enable table-top quantum experiments to simulate complex dynamics, including compressible turbulence, and pulsar glitches in neutron stars that will enabling astrophysical observations to constrain the properties of neutron-rich matter.
    
\section{Introduction}

Computation is regarded as the third pillar of physical science, complementing theoretical and experimental physics.
Each pillar has its unique methodology: theoretical physics relies on mathematical analysis, measurements are the central interest of experimental physics, and numerical modeling is the heart of computational physics.
Many recent breakthroughs, like observing the Higgs boson~\cite{Higgs2012-CMS,Higgs2012-ATLAS} or detecting gravitational waves~\cite{PhysRevLett.116.061102}, would not have been possible without advanced numerical analysis capabilities that adapt algorithmic breakthroughs to evolving hardware.
Here we demonstrate the synergy between theory and computation: advances in linear algebra libraries enable Europe's fastest supercomputer (\gls{LUMI}) to diagonalize matrices of record size, allowing us to simulate turbulent dynamics in quantum systems (superfluids).
We use these simulations to investigate how vortices dissipate energy, driving quantum turbulence in neutron stars and ultracold-atom experiments.

\begin{figure}[t]
	\centering
	\includegraphics[width=0.9\columnwidth]{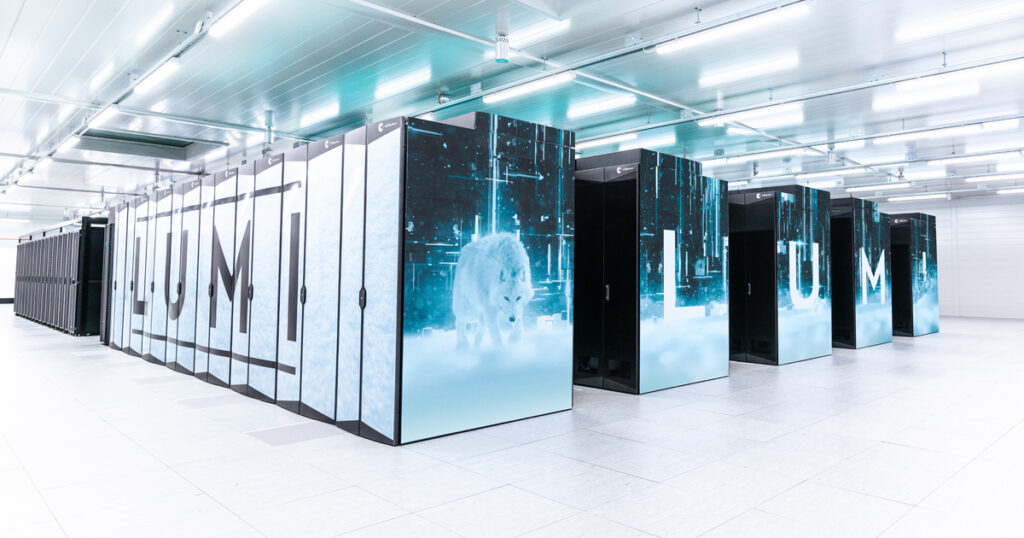}
	\caption[]
  {\gls{LUMI}, \gls{EuroHPC}'s pre-exascale system with \gls{AMD} \mysc{mi250x} \gls{GPU}-accelerated nodes. Each \gls{GPU} node consists of the four \gls{AMD} \mysc{mi250x} \glspl{GPU}, each of which has two \glspl{GCD} being individual \gls{HIP}-programmable devices. Photo copyright Fade Creative.
  }
	\label{fig:lumi}
\end{figure}

As Moore's law bottoms out, using \gls{HPC} effectively becomes a significant challenge.
Current \gls{HPC} systems consist of thousands of interconnected nodes, each comprising dozens of computing cores or multiple hardware accelerators.
Specifically, accelerators like \glspl{GPU} account for most of the computing power on modern platforms.
Leadership supercomputers can compute from \SI{e17}{\gls{FLOPS}} for pre-exascale systems, to \SI{e18}{\gls{FLOPS}} ($=$\SI{1}{E\gls{FLOPS}} or exa-\gls{FLOPS}) for exascale systems.
According to Top 500 list (June 2023) the top three supercomputers are: Frontier (Oak Ridge National Laboratory, \mysc{usa}) with \SI{1.19}{E\gls{FLOPS}}, Supercomputer Fugaku (\gls{RIKEN} Center for Computational Science, Japan) with \SI{0.44}{E\gls{FLOPS}}, and \gls{LUMI} (\gls{EuroHPC}/\gls{CSC}, Finland) with \SI{0.31}{E\gls{FLOPS}}.
Here we use \gls{LUMI} (\cref{fig:lumi}), the fastest European system, to demonstrate
some of its capabilities to advance computational physics.

While this computational potential is enormous, using these \gls{HPC} capabilities requires a highly-tuned software stack capable of dealing with massive parallel and heterogeneous architectures, and core scientific libraries are constantly being adjusted to maximize performance on new hardware.
These include Fast Fourier Transforms, linear algebra routines, libraries for matrix decomposition, random number generators, and solvers for algebraic and differential equations.
These core libraries form the building blocks for the efficient domain-specific scientific packages that enable us to make physics breakthroughs in the domain of quantum mechanics.

Simulating quantum dynamics is one of the hardest challenges for classical computers due to the exponentially large size of a many-body wavefunction.
Even storing the wavefunction for a modest nucleus like tin with $\sim 100$ nucleons would require more bytes than there are atoms in the visible universe.
The techniques of \gls{DFT}~\cite{HK:1964, PhysRev.140.A1133, dreizler1990density} and its time-dependent extension \gls{TDDFT}~\cite{Runge:1984mz, Gross:2012} have revolutionized our ability to study quantum dynamics by replacing the need to store the many-body wavefunction with an energy functional of a handful of densities.
Despite needing to approximate the form of the functional, \gls{TDDFT} has become one of the most successful methods for simulating dynamics in quantum systems.
Its popularity is due to both its accuracy and versatility, providing access to static properties at zero and finite temperatures~\cite{dreizler1990density, Fiolhais2003}, and real-time dynamics~\cite{Gross:2012} for highly complex problems consisting of thousands of particles.
For these reasons, \gls{DFT} is presently the tool of choice for solid-state physics and quantum chemistry~\cite{Zangwill2014, BoostingMaterials, doi:10.1080/00268976.2017.1333644}, and is used extensively in nuclear physics and astrophysics~\cite{doi:10.1080/23746149.2020.1740061, doi:10.1146/annurev-nucl-101918-023608, RevModPhys.88.045004}.

The discovery of high-temperature superconductivity~\cite{Bednorz1986} encouraged extensions of \gls{DFT} to superconducting systems~\cite{PhysRevLett.60.2430, PhysRevLett.73.2915}.
A local formulation was soon developed~\cite{Bulgac2002,Bulgac2007,Bulgac:2012} called
the \gls{SLDA}, which has now been verified and validated against both \gls{QMC} calculations and experiments at the few-percent level for a wide range of systems~\cite{Bulgac:2012, Bulgac:2013b, Bulgac:2016}.
These extensions have brought about new applications for \gls{DFT}: designing superconductors~\cite{Choudhary2022}, simulating nuclear reactions~\cite{Bulgac2016, PhysRevLett.119.042501}, and benchmarking experiments with ultracold Fermi gases~\cite{FGG:2010, Forbes:2012, Bulgac2013, Bulgac2014, Bulgac:2014a, PhysRevLett.120.253002, PhysRevLett.130.023003, PhysRevLett.130.043001}.
Including superfluid correlations, however, comes with a cost that requires super-computing resources.

To properly account for the Pauli exclusion principle, the \gls{SLDA} variant of \gls{DFT} is an orbital-based theory in the spirit of Kohn and Sham~\cite{PhysRev.140.A1133}.
When applied to a normal state, this requires one \term{orbital} per particle.
Thus, to describe $N_p$ particles on a spatial grid $\vect{r}$ requires storing $N_p$ functions.
Extending this to superfluids requires at least twice as many functions, which we call \glspl{qpwf} $\varphi_{n}(\vect{r}, t) = [u_n(\vect{r}, t), v_n(\vect{r}, t)]^T$, since we must represent both particles and holes.
(Including both Hartree and Fock terms in nuclei requires four times as many functions (see e.g.~\cite{Bulgac:2019}).)
The main cost increase, however, is that superfluids allow fractional occupation of the orbitals, and to obtain convergence, one needs significantly more \glspl{qpwf} than there are particles in the system.
As we shall estimate below, if we represent our system on a three-dimensional spatial grid containing $N^3$ points, then we need on the order of $N_{\gls{qpwf}} \approx 0.5 N^3$ \glspl{qpwf} leading to a memory cost of $\approx 0.5 N^6$ complex numbers per state.
For a grid with $N=100$ points in each direction, a single state thus requires $\approx\SI{15}{TB}$.

We consider the simplest type of fermionic superfluid comprising equal populations of two species.
(Think spin-up and spin-down, but in most cold-atom experiments these are different hyperfine states.)
In these systems, the attractive interaction between these two states can be tuned at will using magnetic fields and broad Feshbach resonances to realize fermionic superfluids throughout what is called the \gls{BCS}-\gls{BEC} crossover~\cite{Zwerger:2012}.
For weak inter-particle attractions, the superfluid has a \gls{BCS}-like structure with large Cooper pairs providing long-range order and the associated superfluid flow while slightly modifying the dominant structure of the Fermi sea.
As one increases the interaction strength, the size of these Cooper pairs gets smaller and smaller until they are best described as tightly bound dimers consisting of fermions of each type.
In this \gls{BEC} limit, the dimers behave as bosons, and the system is well described as a \gls{BEC} with the \gls{GPE}~\cref{eq:GPE} where $m_B = 2m$ is now the mass of the dimer, and the dimer density $n_B = n/2$ is half the total fermion density $n$.
In the middle is the so-called \gls{UFG} where the dimers are on the threshold of being bound in the vacuum.
Interestingly, there is no phase-transition separating the strongly and weakly interacting systems, hence the term crossover.

When these systems are sufficiently dilute, all the parameters of the short-range interaction can be become irrelevant except for the $s$-wave scattering length $a$ which describes the size of the bound dimers on the \gls{BEC} side when $a>0$.
At unitarity $a \rightarrow \infty$, and it becomes negative on the \gls{BCS} side where the dimers are unbound.
It is thus convenient to parameterize the equation of state in terms of the dimensionless parameter $\kFa$, where $\kF = (3\pi^2 n)^{1/3}$ is the so-called Fermi momentum.
For reference, the total density $n$, Fermi energy $\varepsilon_F(n)$, and energy density $\mathcal{E}_{FG}(n)$ for this two-component free-Fermi gas are
\begin{subequations}
  \label{eq:FFG}
  \begin{gather}
    \begin{aligned}
      n &= 2\int_{k<\kF}\!\!\!1\frac{\d^3\vect{k}}{(2\pi)^3} = \frac{\kF^3}{3\pi^2}, &
      \varepsilon_{F}(n) &=\frac{\hbar^2\kF^2}{2m},
    \end{aligned} \label{eq:eF}\\
    \mathcal{E}_{FG}(n)
    = 2\int_{k<\kF}\!\!\!\varepsilon_{F}(k)\frac{\d^3\vect{k}}{(2\pi)^3}
    = \frac{\kF^5}{10\pi^2 m}
    = \frac{\hbar^2(3\pi^2)^{5/3}}{10m\pi^2}n^{5/3}.
  \end{gather}
\end{subequations}
At zero temperature ($T=0$), for example, the equation of state throughout the crossover can be expressed as
\begin{gather}
  \mathcal{E}_{\kFa}(n)
  = \xi(\kFa) \mathcal{E}_{FG}(n),
\end{gather}
The dimensionless function $\xi(\kFa)$ characterizes the strength of the interactions, and a major challenge for both theory and experiment has been to determine the universal value at unitarity known as the Bertsch parameter $\xi(\infty) \approx 0.37$~\cite{Forbes:2012}, which combines both experiment~\cite{Ku:2012, Zurn:2013} and \gls{QMC}~\cite{FGG:2010, Carlson:2011} results.
(Note: computing the value of $\xi(\infty)$ was a race between classical and
analog quantum computing using cold atoms.  Establishing the value was one of the first
demonstrated successes of the quantum approach.)
Dynamics in these systems has direct application to ultra-cold atom experiments, but also indirectly to nuclear physics.
In particular, the neutron-neutron scattering length is accidentally large, making the \gls{UFG} an excellent proxy for the dilute neutron matter expected to occur in the crust of neutron stars~\cite{Gezerlis;Carlson:2008-03,Lopez2022,Bulgac2019}.

Simulating quantum turbulence is one of the most complex problem in quantum mechanics.
The phenomenon has been studied intensively using both isotopes of superfluid helium -- bosonic \ce{^4He}, and fermionic \ce{^3He} (see~\cite{Barenghi2001,TSUBOTA2013191} for reviews) -- and more recently in ultracold atomic gases (see~\cite{PhysRevLett.103.045301,PhysRevA.79.043618,Caracanhas2011,Navon2016,Tavares2017,PhysRevResearch.5.043081} and~\cite{Bagnato:2016, Madeira2020} for reviews).
The cold-atom platform has several advantages over liquid helium~\cite{Bulgac:2016}, including access to compressible and hypersonic regimes, superfluid mixtures for studying entrainment~\cite{Hossain:2022a}, and accurate microscopic theories in the form of the \gls{TDSLDA} for fermions, and the \gls{GPE} for bosons (see e.g.~\cite{Tsubota2017}).
We demonstrate here that \gls{HPC} techniques and time-dependent \gls{DFT} frameworks have reached a level of maturity that allows for microscopic simulations of complex phenomena in systems consisting of tens of thousands of superfluid fermions.

\section*{Results}
Here we consider quantum turbulence in an ultra-cold atomic gas of fermions for two cases within  \gls{BEC}-\gls{BCS} crossover:
\begin{enumerate}
\item The strongly coupled \gls{UFG} with $\kFa \rightarrow \infty$.  We compare this to a bosonic (\gls{GPE}) theory for dimers tuned to the \gls{UFG} equation of state;
\item A weakly coupled superfluid in the \gls{BCS} regime with $\kFa = -1.8$.
\end{enumerate}
Through this comparison, we demonstrate the importance of energy dissipation from heating, and its effect on the structure of quantum vortex cores in fermionic systems.

The quantum simulations we perform require two stages of computation: matrix diagonalization to obtain the initial state for evolution, and solving system of millions of coupled \glspl{PDE} to perform the real-time evolution.
Here, we will describe them in the context of an \gls{HPC} implementation on the \gls{LUMI} supercomputer, and benchmark their efficiency on this platform. Before describing the technical aspects of the computation, we introduce a theoretical framework to show the source of the challenges we face.

\subsection*{Theoretical Framework: TDDFT for Bosons}
We start with the simpler problem of describing a bosonic superfluid gas.
If the gas is sufficiently dilute, then the superfluid state can be well described as a \gls{BEC} where all the bosons (dimers in our case) occupy the same condensate wavefunction $\psi_B(\vect{r}, t)$ normalized to the total boson number density $n_B(\vect{r}, t) = \lvert\psi_B(\vect{r}, t)\rvert^2$.
This evolves under a non-linear Schrödinger equation called the \gls{GPE} (see~\cite{Pethick:2002, Pitaevskii:2003}),
\begin{gather}
  \I\hbar e^{\I\eta}\pdiff{\psi_B(\vect{r}, t)}{t} = \underbrace{\left(
      \frac{-\hbar^2\nabla^2}{2m_B} + \mathcal{E}'\bigl(n_B(\vect{r}, t)\bigr)
    \right)}_{\op{h}_B(\vect{r}, t)}
  \psi_B(\vect{r}, t),
  \label{eq:GPE}
\end{gather}
where $m_B$ is the boson mass, $\eta$ is a small phase factor to model dissipation that we tune to better match the natural dissipation in the fermionic simulations~\cite{PhysRevA.105.013304}, and interactions enter through the derivative of the equation of state $\mathcal{E}(n_B)$ that characterizes the energy density as a function of the boson density $n_B$.
This derivative is an effective mean-field chemical potential $\mu = \mathcal{E}'(n_B)$ which repels or attracts bosons depending on the sign and strength of the interaction.
One can include an external potential $V_{\text{ext}}(\vect{r}, t)$ in the single-particle Hamiltonian $\op{h}_B(\vect{r}, t)$, but we do not include one in our simulations here.
Although expressed as a wavefunction, this is equivalent to an orbital-free \gls{DFT}~\cite{PhysRevA.90.043638} of the Hohenberg-Kohn type~\cite{HK:1964}, and \cref{eq:GPE} follows from a principle of stationary action for a generalized Schrödinger field $\psi_B$ with the right-hand-side of \cref{eq:GPE} minimizing the energy functional
\begin{gather}
  E_{\gls{GPE}}[\psi_B] = \int \Bigl(
    \frac{\hbar^2\abs{\vect{\nabla}\psi_B(\vect{r}, t)}^2}{2m_B}
    +
    \mathcal{E}\bigl(\abs{\psi_B(\vect{r}, t)}^2\bigr)
  \Bigr)\d^3{\vect{r}}.
\end{gather}

For modest system sizes, \cref{eq:GPE} can be efficiently solved on small computers, with \gls{HPC} resources being required only for large simulation volumes (see e.g.~\cite{Kobayashi:2021}).
These superfluids demonstrate a wide array of interesting properties, including dissipationless flow past obstacles (with $\eta=0$), and quantized vortices that mediate the energy cascades associated with quantum turbulence in spite of the lack of dissipation~\cite{TSUBOTA2013191,Tsubota2017}.

Although the theory for fermions is much more complicated as we shall describe below, there is a limit which can be well described by a modified version of the \gls{GPE}.
This is the so-called \gls{BEC} limit where two fermionic species have sufficiently strong attraction that they form a gas of tightly bound dimers.
These dimers are bosonic in nature, and can be described by a modified \gls{GPE} like \cref{eq:GPE} with $m_B = 2m$, total density $n = 2n_B = 2\abs{\psi}^2$ (i.e.~$\psi$ describes the dimers), total currents $\vect{j} = 2\Im(\psi\vect{\nabla}\psi^*)$, and a properly tuned equation of state $\mathcal{E}$.
(See~\cite{PhysRevA.90.043638} for details: our modified \gls{GPE} with $\mathcal{E}(n) = \mathcal{E}_{\kFa=\infty}(n)$ is what they call the effective Thomas-Fermi \textsc{etf} model.)
To date, the majority of results for quantum turbulence in ultra-cold atomic gases have been simulated using the \gls{GPE}~\cite{Tsubota2017}.

\subsection*{Theoretical Framework: TDDFT for Fermions}
Unlike bosons, fermions cannot occupy the same state due to the Pauli exclusion principle, and a density functional of the Hohenberg-Kohn type would be highly non-local.
Instead, one uses an orbital \gls{DFT} of the Kohn-Sham type~\cite{PhysRev.140.A1133} where the functional is expressed in terms of a Slater determinant of $N_{\gls{qpwf}}$ single-particle orbitals $\varphi_n(\vect{r}, t) = [u_n(\vect{r}, t), v_n(\vect{r}, t)]^T$, which, as described above, must include at least two components to describe particle-hole excitations in a \gls{BCS} type superfluid.
Each of these states evolves under a single-particle Hamiltonian of the form:
\begin{gather}
  \I\hbar \pdiff{}{t}
  \begin{pmatrix}
    u_n(\vect{r}, t)\\
    v_n(\vect{r}, t)
  \end{pmatrix}
  =
  \underbrace{\begin{pmatrix}
    \op{h}(\vect{r}, t) & \Delta(\vect{r}, t)\\
    \Delta^*(\vect{r}, t) & -\op{h}^*(\vect{r}, t)
  \end{pmatrix}}_{\op{\mathcal{H}}(\vect{r}, t)}  
  \underbrace{\begin{pmatrix}
    u_n(\vect{r}, t)\\
    v_n(\vect{r}, t)
  \end{pmatrix}}_{\varphi_n(\vect{r}, t)},\label{eq:TDSLDA}
\end{gather}
where $\op{h}(\vect{r}, t)$ has a form similar to $\op{h}_B(\vect{r}, t)$ above with second-derivatives in space, and $\Delta(\vect{r}, t)$ is a complex-valued function describing the superfluid correlations.
Together they form the quasiparticle Hamiltonian $\mathcal{H}(\vect{r}, t)$.
The function $\Delta$ plays the role of the order parameter, in analogy to the $\psi_B$ function in the \gls{GPE}.
Unlike in the \gls{GPE}, however, it no longer carries information about the density of the system.
The key to a local \gls{DFT} like the \gls{SLDA} is that $\op{h}(\vect{r}, t)$ and $\Delta(\vect{r}, t)$ depend only on a handful of local densities: the particle density  $n(\vect{r}, t)$, the kinetic density $\tau(\vect{r}, t)$, the current density $\vect{j}(\vect{r}, t)$, and the anomalous density $\nu(\vect{r}, t)$, each of which is computed from the orbitals $\{\varphi_n(\vect{r}, t)\}$ via a reduction.

The precise form of equations of motion~\cref{eq:TDSLDA} will be discussed in the Methods and Materials section below, but follows from minimizing an energy functional of the form
\begin{gather}
  E_{\gls{SLDA}}\bigl[\{\varphi_n(\vect{r}, t)\}\bigr] =
  \int \mathcal{E}_{\gls{SLDA}}\bigl(
  n(\vect{r},t), \tau(\vect{r},t),\vect{j}(\vect{r},t), \nu(\vect{r},t)\bigr)
  \d^3\vect{r}.
\end{gather}
A key property of this system of \glspl{PDE}~\cref{eq:TDSLDA} is that the quasiparticle Hamiltonian $\mathcal{H}$ is the same for all quasiparticle wavefunctions.
This means that at each step of evolution, one needs to communicate only the handful of local densities $\sim$\SI{50}{MB} rather than the complete state.
Furthermore, the single-particle Hamiltonian is unitary, ensuring that the states remain orthonormal throughout the evolution.
Thus, each node can independently perform the evolution of its quasiparticle wavefunctions using local hardware acceleration to compute the derivatives, with minimal communication that requires only an efficient \gls{MPI} reduction.
This is how we solve the second challenge, but to initialize this evolution, we must first obtain a good initial state.

This requires an orthonormal set of quasiparticle wavefunctions $\{\varphi_n(\vect{r})\}$ which solve the self-consistent set of equations minimizing $E_{\gls{SLDA}}[\{\varphi_n\}]$:
\begin{gather}
  \begin{pmatrix}
    \op{h}(\vect{r}) & \Delta(\vect{r})\\
    \Delta^*(\vect{r}) & -\op{h}^*(\vect{r})
  \end{pmatrix}
  \begin{pmatrix}
    u_n(\vect{r})\\
    v_n(\vect{r})
  \end{pmatrix}
  =
  E_n
    \begin{pmatrix}
    u_n(\vect{r})\\
    v_n(\vect{r})
  \end{pmatrix}.
  \label{eq:stHpsi}  
\end{gather}
While the matrix on the left-hand-side can be described efficiently in terms of the densities, the components are formally functions of the orbitals $\op{h}(\varphi_1, \varphi_2, \dots)$ and $\Delta(\varphi_1, \varphi_2, \dots)$.
This Hermitian eigenvalue problem must be solved self-consistently, which we do iteratively through a series of diagonalizations.
In the numerical implementation, the functions $u_{n}(\vect{r})$ and $v_{n}(\vect{r})$ are represented as vectors whose length depends strongly on the geometry and size of the problem.
While some initial states can be computed efficiently -- e.g.\ systems with high degrees of symmetry such as homogeneous matter -- the maximum problem size is generally limited by the technical capabilities of the eigensolver libraries on the chosen \gls{HPC} systems.

\subsection*{Numerical setup and implementation}
To study turbulence, we simulate a periodic volume in space, and use a spectral representation for the quasiparticle wavefunctions on an equally-spaced $N_x \times N_y \times N_z$ Cartesian grid.
In this basis, each quasiparticle wavefunction $\varphi_{n}(\vect{r})$ is represented as a complex vector with $2N_xN_yN_z$ components.

To put the size of our problem in perspective, we focus on a cubic box with $N = N_{x,y,z} = 100$ grid points in each direction.
We define our length scale in terms of the grid-spacing $\Delta x = \Delta y = \Delta z = 1$ so that $V = L^3 = N^3$ and set $\hbar = m = 1$ so that momenta $\vect{p} = \hbar\vect{k}$ is equivalent to the wave-vector.
To compute the kinetic energy in $\op{h}$, we use the \gls{FFT} $\tilde{\varphi}(\vect{k}) = \mathcal{F}\bigl(\varphi(\vect{r})\bigr)$ where the momenta $k = 2\pi n/N$ for $n=\{-N/2,\dots, N/2-1\}$ increase in steps of $\Delta k = 2\pi/N$:
\begin{gather}
  -\frac{1}{2}\nabla^2\varphi_n = \mathcal{F}^{-1}\Bigl(\frac{k^2}{2}\mathcal{F}(\varphi_n)\Bigr).
\end{gather}
This replaces a matrix multiplication with two \glspl{FFT} and a single intermediate vector multiplication by $k^2/2$ which is diagonal in momentum space.
The remaining calculations are local in position space, simply multiplying $\varphi_n$ by various functions of the densities.

\begin{figure}[t]
	\centering
	\includegraphics[width=0.9\columnwidth]{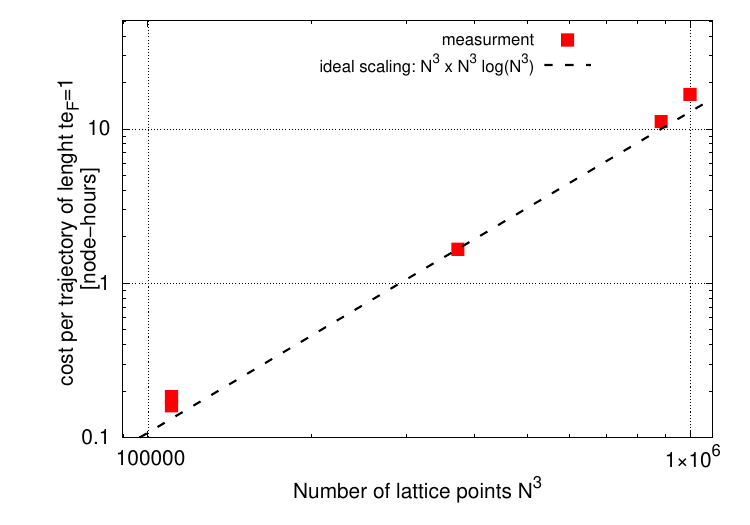}
	\caption[Cost scaling.]{Cost (time-to-solution $\times$ node-count) of the parallel evolution of all quasiparticle wave-functions for a unit time interval (expressed in dimensionless units), which corresponds to 286 integration time steps, as measured on the \gls{LUMI} system.
    The number of nodes was adjusted to fit the problem in the memory available on the nodes, and ranges from 32 in the smallest case to 800 for the largest.
    The dashed line shows the expected ideal cost scaling.}
	\label{fig:td-lumi-scaling}
\end{figure}
These operations can be computed independently and locally on the computation nodes, each of which stores a small fraction of the total set of quasiparticle wavefunctions $\varphi_n$.
The need for \gls{HPC} comes from the large number of these required to adequately represent the problem.
To estimate this, note that, in our units, the maximum momentum represented is $k_{\max} = \pi$, hence the maximum kinetic energy represented is $E_{\max} = k_{\max}^2/2 = \pi^2/2$.
This provides a natural cutoff scale $E_{c} \lesssim E_{\max}$ and we must keep those quasiparticles with energy $E_n < E_c$.
For large $E_n$, the energy is dominated by the kinetic energy, and we can estimate the number of such states by considering the volume $\tfrac{4}{3}\pi k_{\max}^3$ of the sphere $E<E_c$ in momentum space in terms of the volume occupied by each quantum state $(\Delta k)^3$:
\begin{gather}
  N_{\gls{qpwf}} \approx \frac{\frac{4}{3}\pi \overbrace{\pi^3}^{k_{\max}^3}}
                              {\underbrace{(2\pi/N)^3}_{(\Delta k)^3}}
  \approx 0.5 N^3.
\end{gather}
For each state we have two \glspl{PDE} (c.f.~\cref{eq:TDSLDA}). 
This demonstrates the cost of explicitly including the Pauli principle: instead of solving one \gls{PDE} as for bosons, we need to solve many \glspl{PDE}. 
For the case we consider, $N=100$, we will be solving in parallel the corresponding million of \glspl{PDE}, all of them coupled to each other!

\subsection*{Performance of the Time Evolution Algorithm}
The time integration is done with 5\textsuperscript{th}-order multistep \gls{ABM} predictor-corrector method~\cite{Hamming:1973}.
The method requires the evaluation of $\op{h}$ and $\Delta$ twice per time step (predictor and corrector).
The cost per time step thus scales as $N^3 \times N^3\log N^3$ where the first factor accounts for the number of evolved states and the second one for the complexity of the \gls{FFT} that we use to compute the kinetic term.
(We use the \gls{hipFFT} library \myurl{hipfft.readthedocs.io}.) The stability of the method has been studied in~\cite{Jin:2021,PhysRevC.105.044601}. 

\begin{figure}[t]
  \centering
  \includegraphics[clip,trim=0 0 0 30,width=\linewidth]{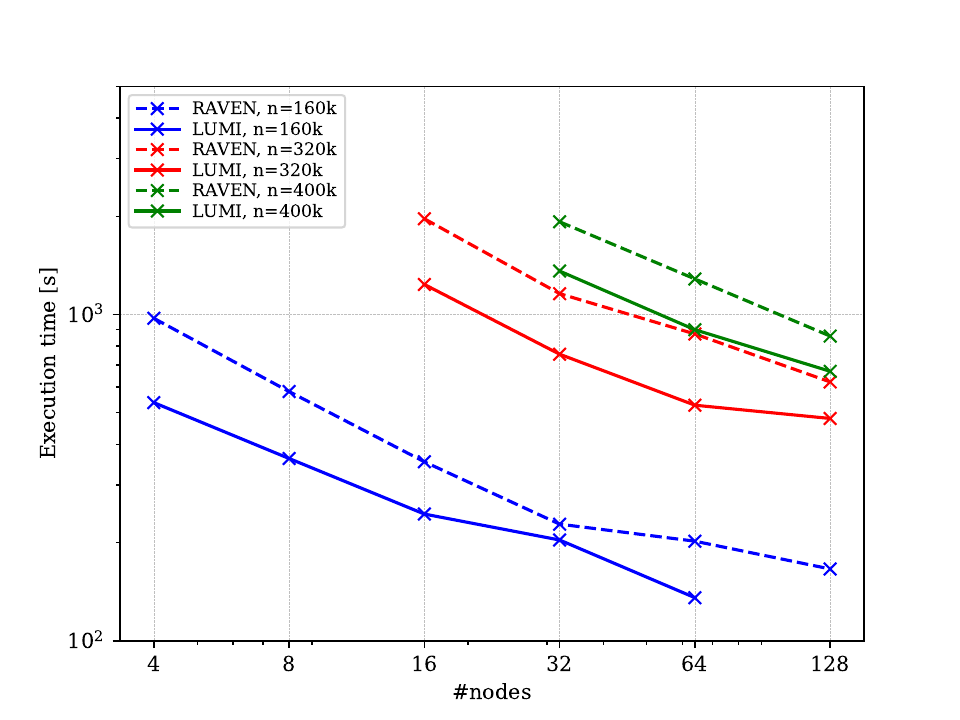}
  \caption[]
  {Strong scaling of the \gls{ELPA} 1-stage solver. 
  Shown is the time-to-solution of the \gls{ELPA} 1-stage solver for real, double-precision computations in a strong-scaling setup for different matrix sizes.
  The solid lines show the results on \gls{LUMI} with 4 \gls{AMD} \gls{MI250X} \glspl{GPU} per node. The dashed lines show the results obtained on the \gls{RAVEN} \gls{HPC} system of the \gls{MPCDF}, with 4 \gls{NVIDIA} \gls{A100} \glspl{GPU} per node.}
  \label{fig:elpa_amd_nividia_compare}
\end{figure}

The open-source \gls{W-SLDA} Toolkit~\cite{WSLDAToolkit} provides a parallel implementation of this time integrator, and is designed to simulate fermionic superfluids with the \gls{TDSLDA} on modern \gls{GPU}-accelerated systems.
In Fig.~\ref{fig:td-lumi-scaling}, we demonstrate the measured cost $C$ (defined as time-to-solution $\times$ node-count) obtained on the \gls{LUMI} system.
Our parallel implementation of~\cref{eq:TDSLDA} exhibits the expected scaling up to the maximum problem size of $N=100$ corresponding to one million \glspl{PDE}.
The main limitation is imposed by the memory requirements: the \gls{ABM} method, while very accurate, require about 10 copies of the state to operate in case of our implementation.
In our largest case ($N=100$) the total memory requirement is about \SI{164}{TB}.

\subsection*{Performance of the Matrix Diagonalization Algorithm}
Generating the initial state is even more costly, as it requires finding a self-consistent solution to \cref{eq:stHpsi} for the complete set of $N_{\gls{qpwf}}$ wavefunctions $\{\varphi_n\}$.
This is done iteratively via a sequence of diagonalizations of an $M\times M$ Hermitian matrix where $M = 2N^3$.
We reduce the cost of iteratively solving~\cref{eq:stHpsi} by using a multi-grid approach.
We fix the domain size and solve the problem on consecutively larger lattices with $N=60$, $80$, and $100$ points in each direction, corresponding to decreasing lattice spacings $\Delta x=1.67$, $1.25$, and $1.0$, respectively.
At each step, we interpolate the converged solution from coarser to finer latices, providing a good initial state to accelerate the iterative algorithm.
In this way, we only need a few iterations to converge on the target $N=100$ lattice.
Even with this tremendous simplification, in the final stage of iterations requires a few dense diagonalization of two million by two million matrices. 

To do this, we use the publicly available \gls{ELPA} library~\cite{ELPA2}, which was designed to efficiently solve dense symmetric or Hermitian standard or generalized \glspl{EVP}, especially with scalability to large core and/or \gls{GPU} counts in mind.
The \gls{ELPA} library was first released in 2010 and has been ported and optimized for all major \gls{HPC} architectures. The \gls{ELPA} library is used in most software packages for electronic-structure theory (See e.g.~\myurl{elpa.mpcdf.mpg.de/ELPA\_USED}.) and clearly outperforms implementations such as \gls{ScaLAPACK}~\cite{ELPA5}.
Recently, in addition to the accelerated version for \gls{NVIDIA} \glspl{GPU}~\cite{ELPA4, ELPA3}, a port to the \gls{AMD} \gls{MI250} \gls{GPU} architecture has been publicly released.
Here we show the first results obtained with this port for \gls{AMD} \glspl{GPU}.

For the standard \gls{EVP}, the \gls{ELPA} library provides two solvers.
The first is a 1-stage solver with three steps: i) transforming the dense matrix into a tridiagonal form, ii) diagonalizing this tridiagonal form, and iii) transforming the eigenvectors back to their original representation.
Alternatively, a 2-stage solver introduces two additional steps: first transforming the dense matrix in a banded matrix, and then transforming the banded matrix into a tridiagonal form;
the subsequent back transformation of the eigenvectors also requires two stages.
For specific algorithmic details, see~\cite{ELPA1, ELPA2}.

\begin{figure}[t]
  \centering
  \includegraphics[clip,trim=0 0 0 30,width=\linewidth]{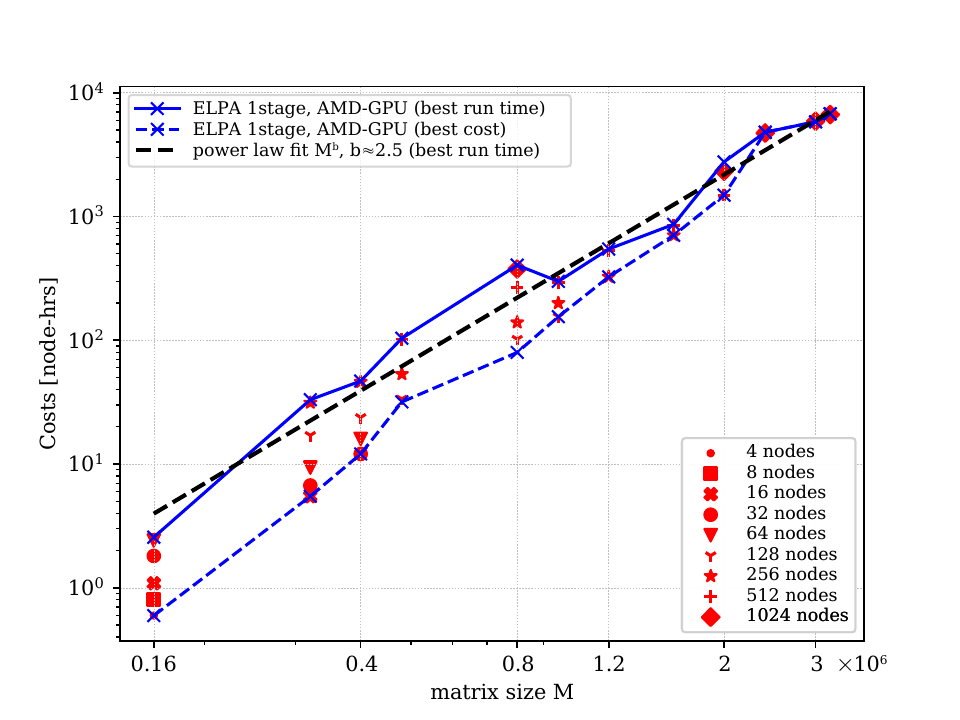}
  \caption[]{%
    Scaling of the \gls{ELPA} 1-stage solver with matrix size.
    Shown is the cost for solution of different eigenvalue problems for different matrix sizes with the \gls{ELPA} 1-stage solver on the \gls{LUMI} \gls{GPU} system.
    The different red symbols indicate the costs (see also \cref{fig:elpa_amd_nividia_compare}) for a fixed matrix size but different node counts.
    The solid blue line represents the fastest time-to-solution (albeit with the highest costs) achieved.
    Note, that the blue line was obtained with measurements on different number of \gls{GPU} nodes.
    Further note that the blue line has been obtained by taking the arithmetic mean of the costs of at least two experiments of a specific matrix size and node count.
    The dashed black line shows a power-law fit to the costs for the best run time.
    The dashed blue line represents the lowest cost (albeit with higher time-to-solution) achieved.
    } 
    \label{fig:elpa_matrix_scaling}
\end{figure}

To initialize the target $N=100$ quantum turbulence problem, we must be able to efficiently diagonalize matrices where the size is of orders of millions by millions.
With access to the \gls{LUMI} supercomputer, we first ensured that the new \gls{ELPA} \gls{AMD} \gls{GPU} version works as expected on large node counts with matrices of this size.
In \cref{fig:elpa_amd_nividia_compare} we compare the run time of the \gls{AMD}
\gls{GPU} version of the \gls{ELPA} library with the run time on \gls{NVIDIA} \gls{A100}
\glspl{GPU} using the \gls{RAVEN} system \myurl{www.mpcdf.mpg.de/services/supercomputing/raven} of the \gls{MPCDF}.
We see that in a direct node-per-node comparison, the solutions of the eigenvalue problems on 4 \gls{MI250X} \glspl{GPU} are in general twice as fast as on 4 \gls{NVIDIA} \gls{A100} \glspl{GPU}, which is in line with the expectations.
Due to limited resources, especially limitations in the maximum job run-time, we could not perform a strong-scaling analysis of the eigenvalue problem for each matrix size -- especially above a linear dimension of one million.
Instead, we had to rather focus on a limited number of experiments and run each eigenvalue solution for a specific matrix size on specific node counts.

This set of runs is shown in~\cref{fig:elpa_matrix_scaling} which shows the costs $C$ to solve a real double-precision dense eigenvalue problem for different matrix sizes.
We include the power-law fit of the $C=M^b$ to the data, representing the scaling behaviour for the best time-to-solution $b\approx 2.5$.
Since this power-law exponents $b$ is still below the theoretical value of 3 (scaling of eigenvalue algorithms is $\mathcal{O}(x^3)$) limit, we have not yet reached the complexity scaling limit for matrix size $M$, at least on this \gls{HPC} system (hardware, compilers, etc.).

Nevertheless, we
have successfully solved dense eigenvalue problems for real, double-precision \glspl{EVP} with linear matrix sizes up to 3.2 million, which is already substantially larger than the problem size required for the target quantum turbulence simulations discussed below.
To our knowledge, this is the largest dense eigenvalue problem ever solved with a direct solver.
While the benchmark of the \gls{ELPA} library was executed for symmetric matrices (double precision), in the production computation, we were working with Hermitian matrices (double complex precision).
The scaling properties in the Hermitian mode are similar, with the complex case taking about twice as long.

\subsection*{Results for Quantum Turbulence}
We have presented the capabilities of leadership supercomputers like \gls{LUMI} to deal with dense matrices and solve nonlinear \glspl{PDE}.
We now combine all these elements together to generate large-scale simulations of turbulent dynamics in ultra-cold Fermi gases.

\begin{figure*}[t]
	\centering
  \labelphantom{fig:td-snapshota}
  \labelphantom{fig:td-snapshotb}
  \labelphantom{fig:td-snapshotc}
  \labelphantom{fig:td-snapshotd}
  \labelphantom{fig:td-snapshote}
	\includegraphics[width=1.0\textwidth]{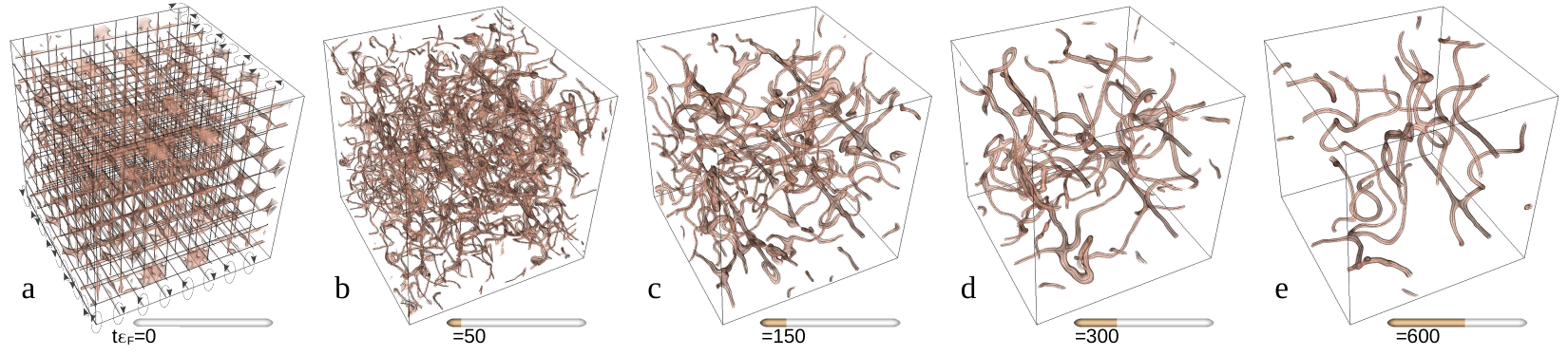}
	\caption[Time evolution of the vortex tangle.]{%
    Selected frames of the time evolution of vortex tangle in the strongly interacting \gls{UFG} ($\kFa\rightarrow\infty$) superfluid gas.
    The lines indicate position of the vortex cores, and isosurfaces are used to visualize their sizes.
    For the simulation we used a periodic lattice with $N^3 = 100^3$ points.
    The full movie is provided in Supplementary Material.}
	\label{fig:td-snapshots}
\end{figure*}
We start our calculations by preparing the initial state at zero temperature ($T=0$) consisting of a regular lattice of imprinted vortices in all three directions, see \cref{fig:td-snapshota}.
The lattice consists of alternately arranged vortices and anti-vortices and the resulting state has zero total angular momentum.
The generation of the initial state amounts to solving the static problem~(\ref{eq:stHpsi}) with the additional constraint imposed on the phase $\theta({\vect{r}})$ of the order parameter $\Delta(\vect{r})=\abs{\Delta(\vect{r})}e^{\I\theta({\vect{r}})}$.
The phase provides a superfluid velocity field $\vect{v}_s(\vect{r}) \propto \vect{\nabla} \theta({\vect{r}})$ consistent with the vortex/anti-vortex lattice with a slight long-wavelength perturbation that destabilizes the vortex lattice leading to a turbulent tangle of vortices as seen in subsequent frames of \cref{fig:td-snapshots}.

We study the strongly-interacting \gls{UFG} ($\kFa = \infty$) with both the full fermionic \gls{TDSLDA} \gls{DFT}, and a simple modified \gls{GPE}-like theory~\cref{eq:GPE} for dimers ($m_B = 2m$) tuned to the \gls{UFG} equation of state~\cite{PhysRevA.90.043638}.
To mimic the natural dissipation of the \gls{TDSLDA}, we add some artificial damping to the \gls{GPE}~\cref{eq:GPE}, considering two values: $\eta = 0.01$ and $\eta = 0.08$ .
The lower value was found to give reasonable qualitative agreement with \gls{UFG} simulations of rotating quantum turbulence~\cite{PhysRevA.105.013304}, while the larger value better matches the flow energy decay seen in the corresponding fermionic simulation.
We also study a less strongly interacting system in the \gls{BCS} regime at an experimentally accessible value of $\kFa=-1.8$.
The number of quasiparticle states extracted from the initial state preparation was $N_{\gls{qpwf}}=\num{582898}$ for the \gls{UFG} and $N_{\gls{qpwf}}=\num{675460}$ for the \gls{BCS} regime.

%
\begin{table}[b]
  \centering
  \begin{ruledtabular}
    \sisetup{table-format=1.1}
    \begin{tabular}{
      S[table-format=6.0]
      S[table-format=-1.1]SSS
      S[table-format=2.0$\xi$]
      S[table-format=5.0$\xi$]
      c}
      \multicolumn{2}{r}{Length scales [$l_F$]:} & {min} & {core} & {separation} & {max} && {method}\\
      {$N$} & {$\kFa$} & {$\Delta x$} & {$\xi$} & {$\lvor(0)$} & 
      {$L_{\text{box}}$} & {$\Lvor(0)$} & \\
      \hline
      22803 & $\infty$ & 1.0 & 1.3 & 7.2 & 99 & 19200 & \gls{GPE}\\
      26790  & $\infty$ & 1.0 & 1.3 & 7.2 & 99 & 19200 & \gls{TDSLDA}\\
      108532 & -1.8 & 1.5 & 2.6 & 14.5 & 150 & 15937 & \gls{TDSLDA} \\
    \end{tabular}
  \end{ruledtabular}
  \caption[Physical parameters in the simulations:]{%
    $N$: total number of particles.
    $\xi=\kF/\pi\Delta$: \gls{BCS} coherence length (typical size of the vortex core).
    $L_{\text{box}}$: size of the simulation domain.
    $a$: scattering length.
    $l_F=\kF^{-1}=[3\pi^2n]^{-1/3}$: inverse of Fermi momentum.
    $\Lvor(0)$: total initial length of vortices.
    $\lvor(0)$: initial mean inter-vortex spacing.
    All lengths are in units of $l_F$.
    The scales are set such ratio $\lvor/\xi\approx 5.5$ is fixed across the runs.
  }
  \label{tab:wslda-params}
\end{table}

\Cref{tab:wslda-params} shows some of the characteristic properties of these initial states.
There are four length-scales of interest.
From smallest to largest, these are:
the Fermi scale $l_F=\kF^{-1}$ is set by the density and is the smallest resolvable scale in the problem;
the \gls{BCS} coherence length $\xi=\kF/\pi\Delta$ describes the size of the cores of the quantum vortices;
the mean inter-vortex distance $\lvor$ describes the vortex density and is the scale at which we initially inject energy for the turbulence;
and the size of the simulation volume $L_{\text{box}}$, which is the largest scale in the problem.

\Cref{fig:td-snapshots} shows the evolution of the \gls{UFG} in our largest \gls{TDSLDA} simulation on \gls{LUMI}.
The initial perturbed vortex lattice (\cref{fig:td-snapshota}) is unstable, and rapidly forms a vortex tangle (\cref{fig:td-snapshotb}) by $t\eF\approx 50$.
The subsequently decay of this tangle (\cref{fig:td-snapshotc,fig:td-snapshotd,fig:td-snapshote}) transfers hydrodynamic energy from the initial scale of the vortex lattice to other length scales.
The bending, crossing, and reconnection of vortices seen in \cref{fig:td-snapshotb,fig:td-snapshotc,fig:td-snapshotd} are the primary mechanisms for quantum turbulence.
Through these mechanisms, hydrodynamic energy can flow from large to small scales, resulting in the emergence of an \term{effective viscosity}~\cite{Babuin2014,PhysRevA.99.043605} even though this is a superfluid.
In compressible fluids, part of this energy is converted into sound~\cite{PhysRevLett.86.1410, PhysRevLett.125.164501} and further into internal excitations, making this cascade irreversible.
There is also some \term{weak wave turbulence}~\cite{Nazarenko:2011,Zakharov:2012} in the phonons (sound waves), but this is a small effect, and not visible in these plots.

To quantify these, we introduce flow and condensation energies.
The former is just the kinetic energy associated with the flow, while the latter estimates the energy in the condensate (Cooper pairs) which uses a simple formula derived in the \gls{BCS} limit~\cite{PhysRevLett.130.023003}:
\begin{align}
  \Eflow(t) &= \int\! \frac{\vect{j}^2(\vect{r},t)}{2n(\vect{r},t)} \d^3\vect{r}, &
  \Econd(t) &= \tfrac{3}{8}\!\int\frac{\abs{\Delta (\vect{r},t)}^2}{\eF (\vect{r},t)} n(\vect{r},t)
  \d^3\vect{r}.
  \label{eq:Eflow}
\end{align}
We compare the evolution of $\Eflow$ with the total length $\Lvor(t)$ of the vortices in \cref{fig:td-decaya,fig:td-decayb}.
When the vortex core is small (we shall call these \term{tight} vortices) as in e.g.\ liquid helium, the flow energy associated with turbulence is dominated by vortices and expected to be proportional to the vortex length.
Our case qualitatively differs from that of liquid helium because we have compression modes (phonons or sound), and the vortex core size is comparable to the inter-particle separation.
Nevertheless, we still see quite a strong correlation between these.

\begin{figure}[tb]
	\centering
 	\labelphantom{fig:td-decaya}
 	\labelphantom{fig:td-decayb}
 	\labelphantom{fig:td-decayc}
 \includegraphics[width=0.72\columnwidth]{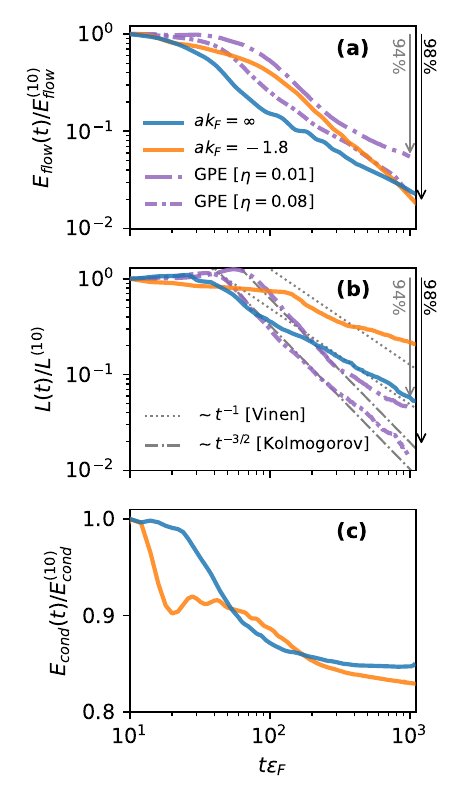}
	\caption[Energy flow.]{%
    Time evolution of: \textbf{(a)} the flow energy $\Eflow$, \textbf{(b)} total vortex length $\Lvor$, and \textbf{(c)} the condensation energy $\Econd$.
    All quantities are normalized with respect to their initial values.
    The solid lines are from the \gls{TDSLDA} calculations on \gls{LUMI} for the \gls{UFG} (dark blue) and \gls{BCS} (light orange) regimes.
    The dashed-dotted lines correspond to the simplified \gls{GPE} model with two different phenomenological dissipation coefficients $\eta\in\{\num{0.01}, \num{0.08}\}$.
    In panel (b), the thin gray lines show the expected slope for $\Lvor(t)$ expected for Vinen (dotted) or Kolmogorov (dashed-dotted) turbulence.
    The arrows indicate where the analyzed quantity drops by \SI{94}{\percent} and \SI{98}{\percent}.
  }
	\label{fig:td-decay}
\end{figure}

The evolution of the total length $\Lvor(t)$ during the free decay is one of the main probes to distinguish the type of turbulence~\cite{Walmsley2014}.
In an incompressible fluid, a random tangle of tight vortices with no large-scale structures in the velocity field is expected to develop what is called \term{Vinen} or \term{ultraquantum} turbulence where the total vortex length decays as $\Lvor \propto t^{-1}$.
On the other hand, if the vortices create large-scale structures -- i.e.\ bundles of coherent vortices -- then one expects the fluid to develop eddies and dynamics that produce \term{Kolomogorov} or \term{quasiclassical} turbulence~\cite{PhysRevLett.109.205304, Baggaley2012} with a characteristic decay of $\Lvor \propto t^{-3/2}$.
In our simulations (\cref{fig:td-decayb}), we see Kolomogorov-like decay in the \gls{GPE} when the vortex density is high $50 \lesssim t\eF \lesssim 300$, but the \gls{TDSLDA} never develops this behaviour, suggesting a fundamental difference in the decay mechanism between bosonic and fermionic simulations.
In this regard, the \gls{TDSLDA} appears to more closely match the decay predicted by Vinen turbulence, but we suspect this is coincidental rather than causal since, as we show below, there are additional dissipation mechanisms present in this case. Note that in this work, we do a comparative study between two methods (\gls{GPE} vs \gls{TDSLDA}) since we have access to data for identical setups. However, the lack of statistics does not allow us to make statements about the precise value of the decay exponent. 

For compressible turbulence, one can use a Helmholtz decomposition to split $\Eflow$ into divergence-free (incompressible/rotational/vortices) and curl-free (compressible/irrotational/phonons) parts $\Eflow=E_{\text{vortices}} + E_{\text{phonons}}$.
(See e.g.~\cite{PhysRevLett.78.3896}.)
The total vortex length $\Lvor$ should be most strongly correlated with the $E_{\text{vortices}}$ contribution, but now vortex reconnections can produce phonons~\cite{PhysRevLett.86.1410, PhysRevLett.125.164501}, further reducing $E_{\text{vortices}}$.
Thus, we expect the decrease in $\Lvor(t)$ to be more pronounced than in $\Eflow(t)$.

To check if this expectation is seen in \cref{fig:td-decay}, we note that, by coincidence, both \gls{TDSLDA} simulations have lost about \SI{98}{\percent} of their total flow energy at $t\eF \approx 1000$, so we use this as a fiducial.
(The phenomenological dissipation $\eta=0.08$ was chosen for one \gls{GPE} to match these results; the other \gls{GPE} simulation with $\eta=0.01$ has a \SI{94}{\percent} loss in $\Eflow$ at this time.)
Here, we find that the \gls{GPE} demonstrates the expected trend for both cases, with a slightly greater drop in $\Lvor$.
The \gls{TDSLDA} simulations, however, have a quantitatively different behaviour with a smaller decrease in $\Lvor$ of \SI{94}{\percent} in the \gls{UFG}, and \SI{80}{\percent} in the \gls{BCS} regime.

\begin{figure}[t]
	\centering
  \labelphantom{fig:st-structurea}
  \labelphantom{fig:st-structureb}
  \labelphantom{fig:st-structurec}
  \labelphantom{fig:st-structured}
  \includegraphics[width=0.72\columnwidth]{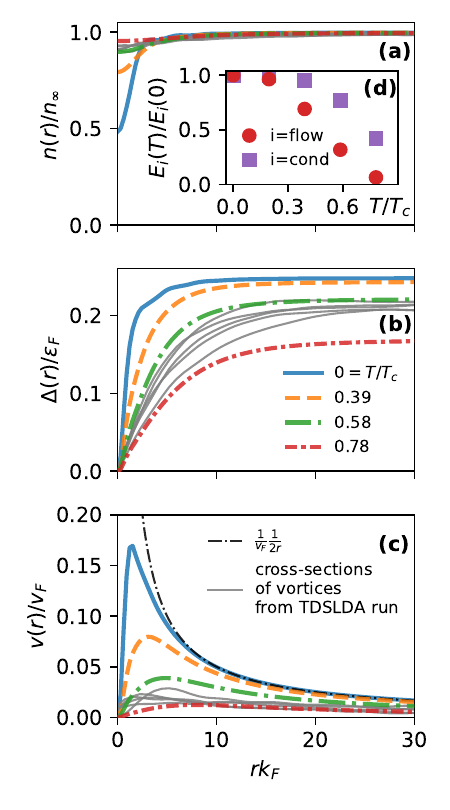}
  \caption[Temperature dependence of the vortex core in the \gls{BCS} regime.]{%
    Radial dependence of the: \textbf{(a)} density $n(r)$, \textbf{(b)} order parameter $\Delta(r)$, and \textbf{(c)} velocity $v(r) = j(r)/n(r)$ for a single straight vortex line at various temperatures in the \gls{BCS} regime ($\kFa=-1.8$).
    The thin gray lines show the corresponding profiles of five randomly selected vortices from the full \gls{TDSLDA} taken at time $t\eF=1000$.
    The temperature is normalized by the critical temperature of superfluid-normal phase transition $T_c$ within the static model: i.e.\ for $T=T_c$ one would have $\Delta(r) = 0$.
    In the lower panel (c) we include the asymptotic form $v(r) = 1/2r$ as a thin dash-dotted line.
    Inset \textbf{(d)}: temperature dependence of $\Eflow$ and $\Econd$ energies for a single vortex.
    Together, these suggest an effective temperature of $T\approx 0.6T_c$ at time $t\eF=1000$.
  }
	\label{fig:st-structure}
\end{figure}

This result might seem to be counterintuitive: in the \gls{BCS} regime we still have many vortices at the end, but they do not generate much flow.
We interpret this as a strong indication that another
  mechanism (absent in the \gls{GPE}) is responsible for reducing $\Eflow$ in the \gls{TDSLDA}.
Noting that the simple correlation $\Lvor \sim E_{\text{vortices}}$ holds only if we do not consider corrections from the internal vortex core structure, we hypothesize that thermalization plays a significant role.
To demonstrate this, in \cref{fig:st-structure} we consider a vortex solution for the \gls{BCS} case, as a function of temperature $T$, which allows us to manipulate the size and structure of the vortex core.
The results are obtained by solving static~\cref{eq:stHpsi} with the constraint that we have a single and straight vortex line. 
Far from the core, the density $n(r)$ has the same behavior, but clearly, the density inside the vortex core is sensitive to the temperature (\cref{fig:st-structurea}).
The order parameter distribution $\Delta(r)$ also indicates that vortices get bigger with the increase of $T$ (\cref{fig:st-structureb}).
Accordingly, the velocity $\vect{v}(r) = \vect{j}(r)/n(r)$, which quantifies the flow energy, is suppressed by the thermal effects (\cref{fig:st-structurec} and inset \cref{fig:st-structured}).
This shows that the structure of the vortex core, which is sensitive to $T$, can affect $\Eflow$.
Specifically, in the inset~\cref{fig:st-structured}, we see that finite temperature significantly reduces both the flow and condensation energy.

This suggests an explanation for the breakdown of the correlation between $\Eflow(t)$ and $\Lvor(t)$: the \gls{TDSLDA} admits an additional dissipation mechanism whereby flow energy is ``thermalized'', altering the flow structure of the vortices.
To explicitly demonstrate that the vortices in the time-dependent runs get hotter, we added to \cref{fig:st-structure} (thin gray lines) cross-sections through five randomly selected vortices from the \gls{BCS} runs at $t\eF=1000$.
As expected for a non-equilibrium state, the profiles of individual vortices have some variability, but all characteristics -- the density profile, the order parameter profile, and the velocity field -- are consistent with the static solutions obtained for a temperature $T/T_c\approx 0.6$.

\begin{figure}[t]
	\centering
  \includegraphics[width=\columnwidth]{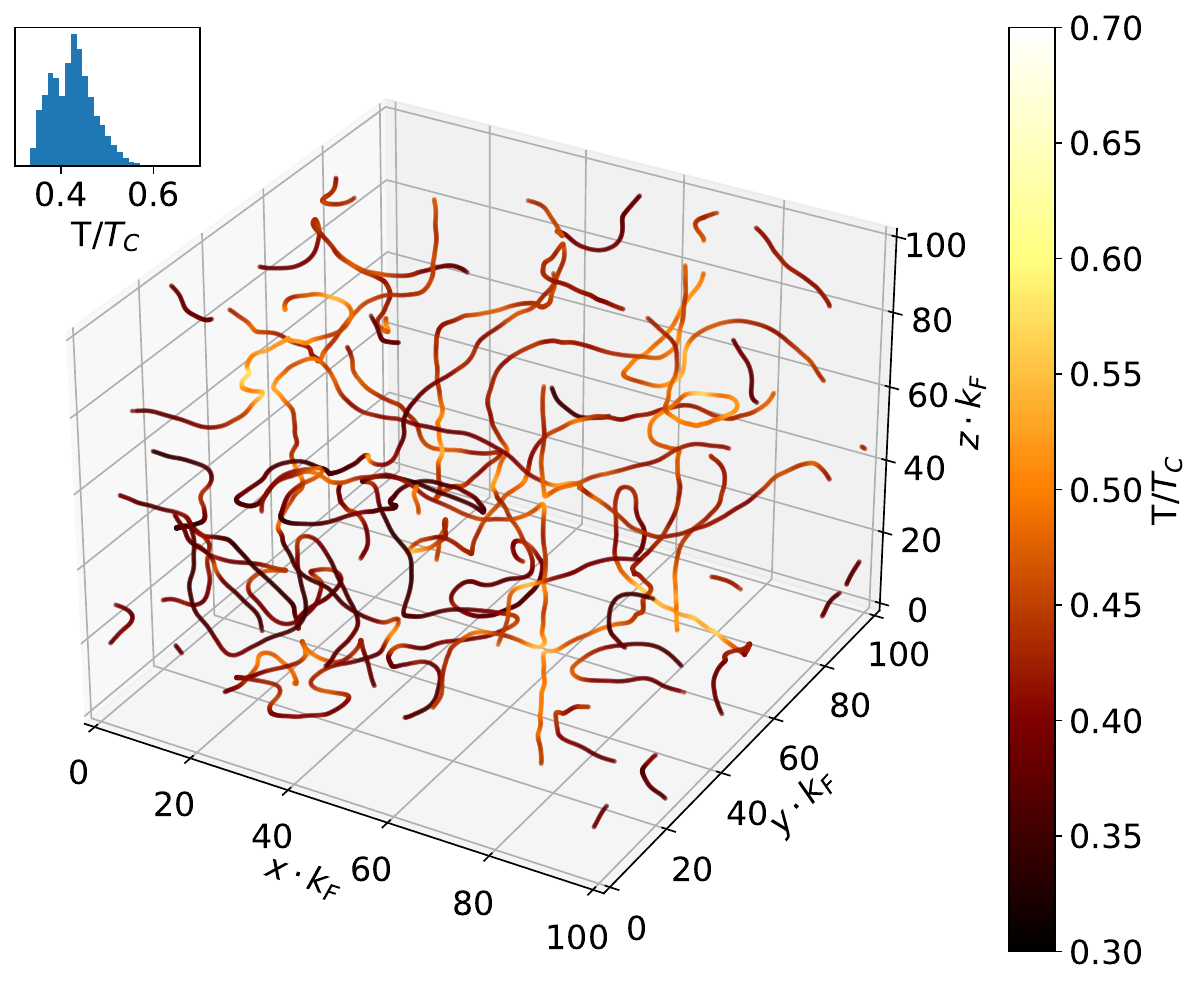}
  \caption[Effective temperature of vortices]{%
    Vortex lines with the color indicating their local effective temperature. The temperature is extracted from the relation between density inside the vortex and the system's temperature for static solutions, see Fig~\ref{fig:st-structure}(a).
      This snapshot is for \gls{UFG} simulation at $t\eF=300$, where the reconnection event occurs and represents the same snapshot as Fig.~\ref{fig:td-snapshotd}.
      The inset shows the histogram of temperatures along the vortex lines.
      The full movie is provided in the Supplementary Material.
      The highest temperatures are generally correlated with events like vortex crossings and reconnections at sites of high curvature, but this energy dissipates, resulting in a uniform effective temperature at later stages of evolution.
  }
	\label{fig:vortex_temperature}
\end{figure}

The temperature dependence of the vortex-core density $n_{\text{core}}$ allows us to use fermionic vortices as a local thermometer.
  Using static \gls{SLDA} simulations of a single vortex, we calibrate $n_{\text{core}}(T)$, (the curve is provided in the Supplementary Material) and then use the density along the vortex lines to demonstrate the thermal evolution of the turbulence in the \gls{TDSLDA} simulations.
  These results are presented in \cref{fig:vortex_temperature}.
  We observe that the effective temperature of vortex lines is higher in regions of higher curvature, especially in regions where reconnections occur.
  This is reminiscent to the heating of wire which is sharply bent back and forth.
  Heating of the vortices represents an additional dissipative mechanism missing in \gls{GPE}-like models.

Heating depletes the condensate, which eventually vanishes at the critical temperature $T_c$ at the superfluid to normal phase transition.
This heating is indicated by the loss of condensation energy as seen in \cref{fig:td-decayc}.
In the \gls{UFG}, the condensate is depleted only at the early stages $t\eF\lesssim 200$, after which it remains relatively constant.
In contrast, in the \gls{BCS} regime, the condensate continues to gradually deplete. 
In the case of the \gls{BCS} vortex solutions, the depletion of the condensate by about \SI{20}{\percent} is for $T\approx 0.6T_c$ (\cref{fig:st-structured}).
This value approximately corresponds to the estimated effective temperature of vortices from the cross-sections. 
It is another signature pointing to the conversion of the flow energy into internal excitations, which effectively heat the system.

\section*{Discussion}
In the \gls{UFG}, the turbulent dynamics resolved by the \gls{TDSLDA} demonstrate qualitative differences when compared with the simplified approach based on a modified \gls{GPE}.
For instance, the \gls{GPE} does not properly account for physics in the cores of vortices: 
Where fermionic vortices have a finite density in the core, vortices in the \gls{GPE} are empty, causing them to move at different speeds.
This can be somewhat compensated for by averaging procedures~\cite{PhysRevA.90.043638}, but feeding this back into the evolution has proved tricky (see \cite{Klimin:2015} for one approach).
It is also insufficient to model the dissipative effects with a single phenomenological parameter $\eta$.
Here we have adjusted $\eta\approx 0.08$ to reasonably match the decay pattern for $\Eflow$ seen in the \gls{TDSLDA}, but this differs from the value $\eta\approx 0.01$ that best fits rotating turbulence~\cite{PhysRevA.105.013304}.
Nevertheless, the \gls{GPE} provides a fast way of gaining some insight into the qualitative effects of superfluid dynamics.

The \gls{TDSLDA} is a parameter-free, self-consistent microscopic theory that naturally captures these effects.
Thus, although more costly computationally, it provides deeper insight into superfluids dynamics than any other currently available techniques.
Here we have demonstrated that the \gls{TDSLDA} breaks the natural correlation between the total flow energy $\Eflow$ and the total vortex length $\Lvor$.
By studying the structure of vortex cores at various temperatures $T$, we provide evidence that this is due to additional energy dissipation and thermalization mechanisms.
Directly comparing the \gls{TDSLDA} with a \gls{GPE}-like theory, we can distinguish this mechanism from dissipation due to vortex bending, crossing, and Kelvin modes, for example, which exist in both theories.
The importance of the fermionic nature of the superfluid is further supported by the result that deviations are stronger in the \gls{BCS} regime than in the \gls{BEC} regime where \gls{GPE}-like theories should be accurate.

Some caveats are in order, and results at finite $T$ in the \gls{SLDA} must be treated with some caution.
In principle, the functional (precisely functions $A$, $B$, and $C$, in~\cref{eq:SLDAform}) should now depend on the dimensionless parameter $k_B T/\varepsilon_F(n)$, but the form of these has not yet been fit due to lack of reliable benchmarks.
Furthermore, it is an open question about how thermalization in strictly $T=0$ dynamical simulations (\cref{fig:td-decay}) should be related to the explicit thermal distribution $f_{T}(E)$ used in static calculations (\cref{fig:st-structure}).
The agreement between these strongly suggests that thermalization is the correct explanation for the break in correlation between $\Eflow(t)$ and $\Lvor(t)$, but additional analysis is needed to conclusively rule out geometric effects, to identify the importance of vortices to the thermalization process~\cite{Silaev:2012a}, etc.

This type of progress requires the development of all exascale \gls{HPC} technology components: computing hardware, optimized scientific libraries for \gls{FFT} (\gls{hipFFT}) and matrix diagonalization (\gls{ELPA}), and advanced high-level scientific application software (the \gls{TDSLDA} via the \gls{W-SLDA} toolkit).
We have demonstrated that, with current technologies, we can diagonalize matrices of order a million by a million, and use \gls{TDDFT} to model fermionic quantum dynamics with \num{e4}-\num{e5} particles.
This is rapidly approaching the scale of typical ultra-cold atom experiments (\num{e5}-\num{e6} particles), allowing us to directly benchmark the \gls{TDSLDA} against experiments.
The new generation of \gls{DFT} packages optimized for \gls{HPC} are extremely flexible, enabling researchers to access not only superfluidity as studied here, but many other fields, including superconductivity, nuclear physics~\cite{RevModPhys.88.045004,Col2020}, nuclear astrophysics, and new opportunities for emerging fields like atomtronics and other quantum technologies~\cite{RevModPhys.94.041001}.  Simulations like these provide access to complex but important microscopic physical processes, and thus provide benchmarks to tune more economical phenomenological models like~\cite{PhysRevA.90.043638, Klimin:2015}, which may ultimately be scaled up to address problems of importance to fundamental physics in nuclear astrophysics.
(E.g., the \gls{UFG} we study here is a good model for the neutron superfluid in the
crust of neutron stars, which is likely responsible for pulsar glitches.)

To maximize the scientific impact of these publicly-funded resources, we have open-sourced our codes~\cite{WSLDAToolkit}, and release the raw data generated by these simulations~\cite{Zenodo} so that others can reproduce our analysis, and use these results for further benchmarking other models, thereby advancing the pace of science.
We hope that other groups will follow our example, making their codes and data available for the community.
This step is needed in order to maximize the amount of knowledge extracted from data obtained by expensive \glspl{HPC} systems, and share research opportunities with groups that do not have direct access to them.

\section{Materials and Methods}\label{sec:Materials-and-Methods}
\subsection*{LUMI specification}
\glsreset{GCD}

\Gls{LUMI} is one of the three \gls{EuroHPC} pre-exascale supercomputers, along with Leonardo and Marenostrum 5.
It is the only major European \gls{HPC} system equipped with \gls{AMD} \gls{GPU} technology.
The numerical results described in this work were collected from the \gls{LUMI} pilot access phase.
The \gls{LUMI} system has a peak performance of nearly \SI{500}{P\gls{FLOPS}}, most of which is delivered by 2560 \gls{GPU} nodes.
Each of these nodes includes a single \gls{AMD} \mysc{epyc} \gls{CPU} with 64 cores and four \gls{AMD} \gls{MI250X} \glspl{GPU}.
The \gls{AMD} \gls{MI250X} has a peak performance of \SI{53}{T\gls{FLOPS}} in double-precision arithmetic.
The \gls{AMD} \gls{MI250X} \gls{GPU} package consists of two independent devices, called \glspl{GCD}.
Each of these \glspl{GCD} has 110 compute units, and \SI{64}{GB} of \gls{HBM2} which can be accessed at a peak rate of \SI{1.6}{TB/s}.
The two \glspl{GCD} in the \gls{MI250X} package are connected by an in-package communication interface with a peak bidirectional bandwidth of up to \SI{400}{GB/s}.
Devices on different packages are linked with either a single or double communication link with a peak bidirectional bandwidth of \SI{100}{GB/s} and \SI{200}{GB/s}, respectively.
Each \gls{GPU} package is directly connected to the Slingshot network providing up to $2\times$\SI{50}{GB/s} peak bandwidth. 
The \gls{AMD} \gls{MI250} \gls{GPU} family is based on the 2\textsuperscript{n}d generation \gls{AMD} \gls{CDNA} ("Compute DNA") architecture which uses \gls{AMD} \gls{ROCm} (Radeon Open Compute) development stack.
\Gls{ROCm} is an open-source collection of drivers, development tools and \glspl{API} for \gls{GPU} programming from the low-level kernel to end-user libraries.
Device kernels are programmable with the \gls{HIP} \gls{GPU} programming language extension.
The \gls{HIP} extension also provides a runtime platform, numerical libraries (including the \gls{hipFFT}), and porting tools.

\subsection*{Density functional theory for superfluid fermions}
We consider here a system with equal densities of two types of fermions interacting with short-range interactions.  In second-quantized notation,
\begin{multline}
  \op{H} = \int\d^3\vect{r}\;
  \frac{
    \op{a}^\dagger(\vect{r})\op{p}^2\op{a}(\vect{r})
    +
    \op{b}^\dagger(\vect{r})\op{p}^2\op{b}(\vect{r})
  }{2m}\\
  +
  \int\d^3\vect{r}\d^3\vect{r}'\;
  V(\lvert\vect{r} - \vect{r}'\rvert)\op{n}_{a}(\vect{r})\op{n}_{b}(\vect{r}'),
  \label{eq:quantum}
\end{multline}
where $\op{n}_{a} = \op{a}^\dagger\op{a}$ and $\op{n}_{b} = \op{b}^\dagger\op{b}$ are the number operators for the two species, expressed in terms of annihilation $\op{a}/\op{b}$ and creation $\op{a}^\dagger/\op{b}^\dagger$ operators that satisfy anti-commutation relations. 
We consider the limit where $V(r)$ is short-ranged, so that the interaction can be completely described by a single parameter, the $s$-wave scattering length $a$.

In general, the state of such a system with $N$ particles must be described by a many-body wavefunction $\psi(\vect{r}_1, \vect{r}_2, \dots, \vect{r}_N)$ which requires an exponential amount of information, but \gls{DFT} allows us to reduce this description to an effective theory for the total density $n(\vect{r}) = \braket{\op{n}_a(\vect{r})} + \braket{\op{n}_{b}(\vect{r})}$ or states which can be expressed as a Slater determinant -- an anti-symmetrized product of single-particle states.
Treated as a variational problem of finding the best single--Slater-determinant state (with some care required to express the zero-range limit), one derives a \gls{BCS}-like ansatz in what is commonly referred to as \gls{HFB} theory or \gls{BdG} theory for superconductivity. (See e.g.~\cite{Pethick:2002, Pitaevskii:2003, Schunck:2019}.)
These theories qualitatively capture properties of the system \cref{eq:quantum}, but quantitatively fail in regions of interest like the \gls{UFG}.
The \gls{SLDA} has a similar mathematical form, but is constructed from a different philosophy -- that of \glsreset{DFT}\gls{DFT}.

The \gls{TDDFT} equations follow from a condition of stationary action
\begin{gather}
  \delta S = 0, \qquad
  S=\int_{t_0}^{t_1} \left(
    \Bigl\langle \Omega(t) \Big | \I\hbar\diff{}{t} \Big | \Omega(t) \Bigr\rangle - E(t)
  \right) \d{t},
  \label{eqn:S}
\end{gather}
where $\ket{\Omega(t)}$ is quasiparticle vacuum at time $t$ and $E(t)$ is the total energy.
The key to all \gls{DFT} approaches is the existence theorem due to Hohenberg and Kohn~\cite{HK:1964}, with extension to the time-dependent cases by Runge and Gross~\cite{Runge:1984mz}, that for any given system, $E(t)$ can be expressed as
\begin{gather}
  E(t) = E_{\text{int}}[n(\vect{r},t)] + \int V_{\text{ext}}(\vect{r}, t)n(\vect{r},t)\d^{3}\vect{r}
\end{gather}
where $E_{\text{int}}[n]$ is a universal functional and $V_{\text{ext}}(\vect{r}, t)$ is the external potential (which we set to zero here).
Unfortunately, no prescription for finding $E_{\text{int}}[n]$ is known, and it is likely extremely complicated and non-local, even for non-interacting fermions.
Instead, Kohn and Sham~\cite{PhysRev.140.A1133} derived an equivalent but alternate formulation in terms of an energy functional of a Slater determinant $\ket{\Omega(t)}$ of single-particle orbitals that allows for an exact local formulation for non-interacting fermions.
By including both particles and holes, and with an appropriate regularization procedure, this was generalized for superfluids~\cite{Bulgac2002, Bulgac2007, Bulgac:2012, PhysRevA.106.013306} in a form called the \gls{SLDA} that we now describe.
Unlike the \gls{BdG} equations, we can now tune the parameters of the theory to match experiment and ab-initio \gls{QMC} calculations.
We lose any notion of a variational bound, but obtain instead a theory accurate to the few-percent level for a wide range of systems~\cite{Bulgac:2012, Bulgac:2013b, Bulgac:2016}.

The energy function for the \gls{SLDA} can be expressed as an integral of four local ``densities'' -- each of which is a function of the orbitals in the Slater-determinant state $\ket{\Omega(t)}$:
\begin{gather}
  \label{eq:local-functional}
	E(t) = \int\d^{3}\vect{r}\;
  \mathcal{E}[n(\vect{r},t), \tau(\vect{r},t),\vect{j}(\vect{r},t), \nu(\vect{r},t)].
\end{gather}
At $T=0$, this system has only two length-scales: the inverse Fermi momentum $\kF^{-1}$ and the $s$-wave scattering length $a$.
The \gls{SLDA} functional is thus constrained by dimensional arguments (see \cref{eq:FFG}) to have the following form (in units where $\hbar=m=k_B=1$):
\begin{gather}
  \mathcal{E} =A(\kFa)
    \frac{\tau}{2}
  +B(\kFa)\frac{3}{5}n\eF
  +\frac{C(\kFa)}{n^{1/3}}|\nu|^2
  +[1-A(\kFa)]\frac{\vect{j}^2}{2n},\label{eq:SLDAform}
\end{gather}
where $A$, $B$, and $C$ are dimensionless universal functions.
The first term defines the kinetic energy, the second term (missing in \gls{BdG} theory) describes the Hartree energy, the third term accounts for the energy gain due to pairing correlations, and the last term is required to restore Galilean covariance.

By appropriately choosing the universal functions, one can describe the entire \gls{BEC}-\gls{BCS} crossover, including the weakly-interaction \gls{BCS} limit $\kFa \rightarrow 0^{-}$, and the \gls{UFG} $\abs{\kFa}\rightarrow \infty$.
The latter is especially simple because $A$, $B$, and $C$, are just numbers.
Precisely, the functions are constructed in such way as to ensure correct reproductions of selected properties of uniform Fermi gas at a given value of the interactions parameter $\kFa$. 
These properties are: equations of state $\xi(\kFa) = \mathcal{E}/\mathcal{E}_{FG}(n)$, strength of the pairing correlations $\Delta(\kFa)/\varepsilon_F(n)$, and the quasiparticle effective mass $m^*(\kFa)=m/A(\kFa)$.
All of these quantities are accessible from \gls{QMC} calculations.
For more details related to the construction of the functional, see~\cite{Bulgac2002, Bulgac2007, Bulgac:2012, Bulgac:2013b, FGG:2010, Bulgac:2016, PhysRevA.106.013306, Bulgac:2019} and the source code for the reference implementation.
In the calculations presented here, we have assumed that the effective mass $m^*=m$, so $A(\kFa) = 1$.
This is a physically reasonable approximation that simplifies the functional slightly since the last term of~\cref{eq:SLDAform} vanishes.

The local densities entering the functional are:
\begin{align}
  n(\vect{r},t)
  &= \braket{\Omega(t)|
    \op{a}^\dagger(\vect{r})\op{a}(\vect{r})
    + \op{b}^\dagger(\vect{r})\op{b}(\vect{r})|\Omega(t)}, \tag{total}\\
  \tau(\vect{r},t)
  &= \braket{\Omega(t)|
    \vect{\nabla}\op{a}^\dagger(\vect{r})\cdot
    \vect{\nabla}\op{a}(\vect{r})
    + \vect{\nabla}\op{b}^\dagger(\vect{r})\cdot\vect{\nabla}\op{b}(\vect{r})|\Omega(t)},
    \tag{kinetic}\\
  \vect{j}(\vect{r},t)
  &= \Im\braket{\Omega(t)|
    \op{a}^\dagger(\vect{r})\cdot
    \vect{\nabla}\op{a}(\vect{r})
    + \op{b}^\dagger(\vect{r})\cdot\vect{\nabla}\op{b}(\vect{r})|\Omega(t)},
      \tag{current}\\
  \nu(\vect{r},t)
  &= \braket{\Omega(t)|\op{a}(\vect{r})\op{b}(\vect{r})|\Omega(t)}.
      \tag{anomalous}
\end{align}
After varying the functional, we will obtain a matrix equation~\cref{eq:TDSLDA} (derived below) with eigenvalues $E_n$ and two-component eigenstates $\varphi_{n}(\vect{r},t)=[u_{n}(\vect{r},t), v_{n}(\vect{r},t)]^T$.
In terms of these, the densities are:
\begin{subequations}
  \label{eq:densities2}
  \begin{align}
    n(\vect{r}, t)
    &= 2\!\!\!\!\!\!\sum_{0<E_n<E_c}\!\!\left(
      \abs{v_{n}}^2 f_{T}(-E_n)+\abs{u_{n}}^2 f_{T}(E_n)
      \right), \label{eq:n}
    \\
    \tau(\vect{r}, t)
    &= 2\!\!\!\!\!\!\sum_{0<E_n<E_c}\!\!\left(
      \abs{\vect{\nabla}v_{n}}^2 f_{T}(-E_n)
      + \abs{\vect{\nabla}u_{n}}^2 f_{T}(E_n)
      \right), \label{eq:tau}
    \\
    \vect{j}(\vect{r}, t)
    &= 2\!\!\!\!\!\!\sum_{0<E_n<E_c}\!\!\!\!\Im\bigl(
         (v_{n}\vect{\nabla} v_{n}^*) f_{T}(-E_n)
         -  (u_{n}\vect{\nabla} u_{n}^*) f_{T}(E_n)
      \bigr), \label{eq:j}
    \\
    \nu(\vect{r}, t)
    &= \phantom{2}\!\!\!\!\!\!\sum_{0<E_n<E_c}\!\!\!
      u_{n}v_{n}^{*}
      \bigl(f_{T}(-E_n)-f_{T}(E_n)\bigr),
  \end{align}
\end{subequations}
where we have suppressed the space-time arguments on the components $u_n$ and $v_n$ to save space, and we have introduced the thermal distribution function (Fermi distribution)
\begin{gather}
  f_{T}(E) = \frac{1}{1 + \exp(E/T)},
\end{gather}
which allows us to approximate finite temperatures $T$.
We used the finite-temperature variant only for understanding the structure of the vortex core as presented in~\cref{fig:st-structure}; the time-dependent runs were executed at $T=0$.

Note that all states up to a specified energy cutoff $E_c = \hbar^2k_c^2/2m$ contribute to the densities: This is the main difference between the \gls{DFT} for superfluid systems and the original formulation of Kohn and Sham which only keeps states up to the Fermi surface.
For superfluids, the single-particle state $\ket{\Omega(t)}$ has a coherent phase and hence does not have a well-defined particle number.
Instead, the \gls{BCS} particle-hole correlations (Cooper pairs) allow fractional occupation of states with larger energy.
The cutoff $E_c$ is essential for a local formulation with short-range interactions because the kinetic $\tau(E_c)$ and anomalous $\nu(E_c)$ densities are linearly divergent in such a state

While both $\tau$ and $\nu$ diverge linearly $\propto k_c$, the combination $\hbar^2\tau/2m - \Delta^\dagger\nu$ is finite where $\Delta = g_c\nu$ is the finite pairing gap.
To improve convergence~\cite{Bulgac2002}, we choose a scale $k_0 \approx \kF$ and use 
\begin{gather*}
  \Lambda_c 
  =
  \frac{m}{\hbar^2}\frac{k_c}{2\pi^2}\left\{
    1 - \frac{k_0}{2k_c}\ln\frac{k_c+k_0}{k_c-k_0}
  \right\}
\end{gather*}
so that the functional is specified by the density dependence of the finite combination $A/g_c + \Lambda_c$.
This finite quantity is proportional to the inverse scattering length $1/a$ in the standard variational formulation of \gls{BdG} theory, but has additional dependence on $\kFa$ in the \gls{SLDA}.
The regularization scheme is constructed to assure independence of the observables with respect to the energy cutoff $E_c$ (assuming it is large enough to encapsulate meaningful states) when considering static problems.
  In the context of time-dependent problems, another factor needs to be considered: the total energy must be conserved.
  Formally, the \gls{TDSLDA} equations~(\ref{eq:TDSLDA}) are conservative only in $E_c\rightarrow\infty$ limit~\cite{Magierski2018}.
  For this reason, in the computation, we use the cutoff scale $E_c=k_{\max}^2/2$ defined through the maximum value of momentum $k_{\max}=\pi$ resolved by our spatial grid.
  In our simulations, the relative change in of the total energy $ \abs{E(t)-E(0)}/E(0)$ does not exceed a fraction of a percent.
See~\cite{Bulgac2002, Bulgac2007, Bulgac:2012, Bulgac:2013b, FGG:2010, Bulgac:2016, PhysRevA.106.013306, Bulgac:2019} and the source code for further details.

The equations of motion that emerge from the stationarity condition \cref{eqn:S} have the forms \cref{eq:TDSLDA,eq:stHpsi} discussed in the results above.
The single particle Hamiltonian $\op{h}$ and pairing potential $\Delta$ are defined through functional derivatives of the energy density
\begin{subequations}
  \begin{align}
    \op{h} &= -\vect{\nabla}\frac{\delta E}{\delta \tau }\vect{\nabla}
      + \frac{\delta E}{\delta n}
      -\frac{\I}{2}\left(
        \frac{\delta E}{\delta \vect{j}}\vect{\nabla}
        + \vect{\nabla}\frac{\delta E}{\delta \vect{j}}
      \right)
    - \mu, \\
    \Delta &= -\frac{\delta E}{\delta \nu^*}
             = \underbrace{-\frac{C}{n^{1/3}}}_{g_c}\nu.
  \end{align}
\end{subequations}
They depend on densities, which in turn depend on the quasiparticle orbitals $[u_{n}(\vect{r}), v_{n}(\vect{r})]^T$.
The single particle Hamiltonian $\op{h}$ includes a shift by the value of the chemical potential $\mu$.
This controls the particle number in the static solution for the initial state \cref{eq:stHpsi}.
It is formally irrelevant for the time-evolution \cref{eq:TDSLDA}, but helps improve numerical convergence by minimizing the evolution of the global phase.
The regularized coupling function $g_c$ defines the order parameter $\Delta$, ensuring it remains finite.

\subsection*{Initial state preparation and vortex lines detection}
To generate the quantum turbulence, we start with a regular alternating array of interleaved vortices and anti-vortices in all three directions.
(See \cref{fig:td-snapshota}.)
Using the Biot-Savart law, including image vortices from the periodic box, we obtain the phase profile $\theta_0(\vect{r})$ for this periodic array with a superfluid velocity field $\vect{v}_s(\vect{r}) \propto \vect{\nabla} \theta_0({\vect{r}})$ consistent with the vortex/anti-vortex lattice.
On this periodic phase profile, we add a few small low-frequency Fourier components $\theta(\vect{r}) = \theta_0(\vect{r}) + a_0\sum_{n=0}^{N_v} c_n\cos(\vect{k}_n \cdot\vect{r})$, where $N_v$ is the number of vortices in each direction.
The coefficients and frequencies are $c_n = (-1)^n / (2n + 1)^2$, and $k_n^i = (2n + 1) (2\pi)/L_\mathrm{box}^i$.
The magnitude of the perturbation $a_0$ is adjusted so that the additional large-scale flow increases $E_{\text{flow}}$ by 5\% compared with the vortex lattice (as computed within the \gls{GPE}).
This phase profile is then held fixed in $\Delta(\vect{r}) = \abs{\Delta(\vect{r})}e^{\I\theta(\vect{r})}$, and the iterative solution for the magnitude of the gap $\abs{\Delta}$, density, etc.\ is found using \cref{eq:stHpsi} as described above, which requires the improved \gls{ELPA} diagonalization routines.

To compute the total length of $L$, we need first to identify the position of vortices from numerical data. For this, we search in $\Delta(\vect{r})$ fields for points around which its phase winds. This also implies that, at this point, the order parameter vanishes. We interpolate grid data using \gls{DVR} to identify such points with subgrid resolution~\cite{PhysRevC.87.051301}. To connect points into lines we use pseudovorticity $\bm{\omega}(\bm{r})=\bm{\nabla}\times\bm{j}(\bm{r})$, which should point along the line. A detailed description of the algorithm is given in~\cite{PhysRevA.105.013304}.

\backmatter

\section*{Acknowledgments}
    The \gls{MPCDF}, and especially A.~Marek, is grateful for the fruitful discussions and support of
    \gls{AMD} for optimizing the \gls{ELPA} library on \gls{AMD} \glspl{GPU}.
    The authors are grateful to the support team of the \gls{CSC} data center and their help with using the \gls{LUMI} \gls{HPC} system.
    We thank P.~Magierski for inspiring discussions, and C.~Barenghi for suggesting additional tests.
    This work was supported by the Polish National Science Center (NCN) under Contract No. UMO-2017/26/E/ST3/00428, and later stages of this work were supported under Contract No. UMO-2022/45/B/ST2/00358.    
    The development of the \gls{ELPA} library is financially supported by the NOMAD Centre of Excellence (European Commission grant agreement ID 951786).
        S.S.\ and M.M.F.\ acknowledge \supportfromNSFgrant[PHY]{2012190}. 

\section*{Author contributions}
    The work was conceived by G.W., and M.M.F.
    The \gls{TDSLDA} calculations were done by G.W.
    The \gls{GPE} calculations were done by S.S.
    Porting of the \gls{W-SLDA} Toolkit to \gls{AMD} was done by G.W., and M.S.
    Porting and optimizing of \gls{ELPA} to \gls{AMD} was done by A.M.
    Benchmarking \gls{ELPA} on \gls{LUMI} was done by A.M.
    Data analysis was done by G.W., S.S., and M.M.F.
    Support of computation processes on \gls{LUMI} was done by M.S.
    All authors contributed to the manuscript writing.

\section*{Data and materials availability}
      Data underlying the study is available via Zenodo repository~\cite{Zenodo}.
      
\bibliography{macros,biblio}

\section*{Supplementary Materials}
\subsection{Raw Data}
We provide raw data through Zenodo repository~\cite{Zenodo}. 
The datasets include:
\begin{enumerate}
    \item dataset with results for BCS run ($a\kF=-1.8$) obtained with TDSLDA method (\verb|bcs| folder).
    \item results of static calculations for single vortex line for BCS run ($a\kF=-1.8$) obtained with SLDA method (\verb|bcs_static_vortex| folder).
    \item dataset with results for UFG run ($a\kF=\infty$) obtained with TDSLDA method (\verb|ufg| folder). 
    \item dataset with results for UFG run ($a\kF=\infty$) obtained with GPE method. The results are provided for two dissipation coefficients $\eta=0.01$ and $0.08$ (\verb|ufg_gpe| folder)
\end{enumerate}
The files contain the following information:
\begin{itemize}
    \item Time evolution of the density $n(\bm{r},t)$, the current $\bm{j}(\bm{r},t)$ and the order parameter $\Delta(\bm{r}, t)$. 
    \item Positions of extracted vortex lines from time-dependent runs.
    \item{Reproducibility packs for TDSLDA runs. They contain the full information about the settings and the computation process.}
    \item Static solutions for the single vortex in the BCS regime ($a\kF=-1.8$). 
\end{itemize}
In the repository, there is \verb|README.md| file that provides information about used data formats. There are also provided example scripts/codes demonstrating how to read data in python and C. 

\subsection{List of Movies}
Below, we provide the list of accompanying movies (in mp4 format). 
3D views were created by VisIt software~\cite{HPV:VisIt}.
The visualizations presented the volume distribution of the order parameter and lines indicating the vortex cores' location.  
All movies are also accessible on YouTube. 

\begin{description}[topsep=-1pt]
    \setlength{\itemsep}{0pt}%
    \setlength{\parskip}{0pt}%
\item[Supplementary movie 1] Run for the unitary Fermi gas ($a\kF=\infty$) by means of TDSLDA.\\
             YouTube: \myurl{youtu.be/hOLPmPVQ4xo}
\item[Supplementary movie 2] Run for the BCS regime ($a\kF=-1.8$) by means of TDSLDA.\\
             YouTube: \myurl{youtu.be/OpIDOPdaPKM}
\item[Supplementary movie 3] Run for the unitary Fermi gas ($a\kF=\infty$) by means of GPE, with the dissipation coefficient $\eta=0.01$.\\
             YouTube: \myurl{youtu.be/3Dux6PHD4e4}
\item[Supplementary movie 4] Run for the unitary Fermi gas ($a\kF=\infty$) by means of GPE, with the dissipation coefficient $\eta=0.081$.\\
             YouTube: \myurl{youtu.be/7FXl-NlAq20}
\item[Supplementary movie 5] Dynamics of vortex lines and their local effective temperatures for the unitary Fermi gas ($a\kF=\infty$).\\
             YouTube: \myurl{youtu.be/Wd\_gEUZcbu4}
\item[Supplementary movie 6] Dynamics of vortex lines and their local effective temperatures for the BCS regime ($a\kF=-1.8$).\\
             YouTube: \myurl{youtu.be/zGSaVhT3e74}
\end{description}

\subsection{Thermometer calibration curve}

\begin{figure}[t]
	\centering
  \includegraphics[width=\columnwidth]{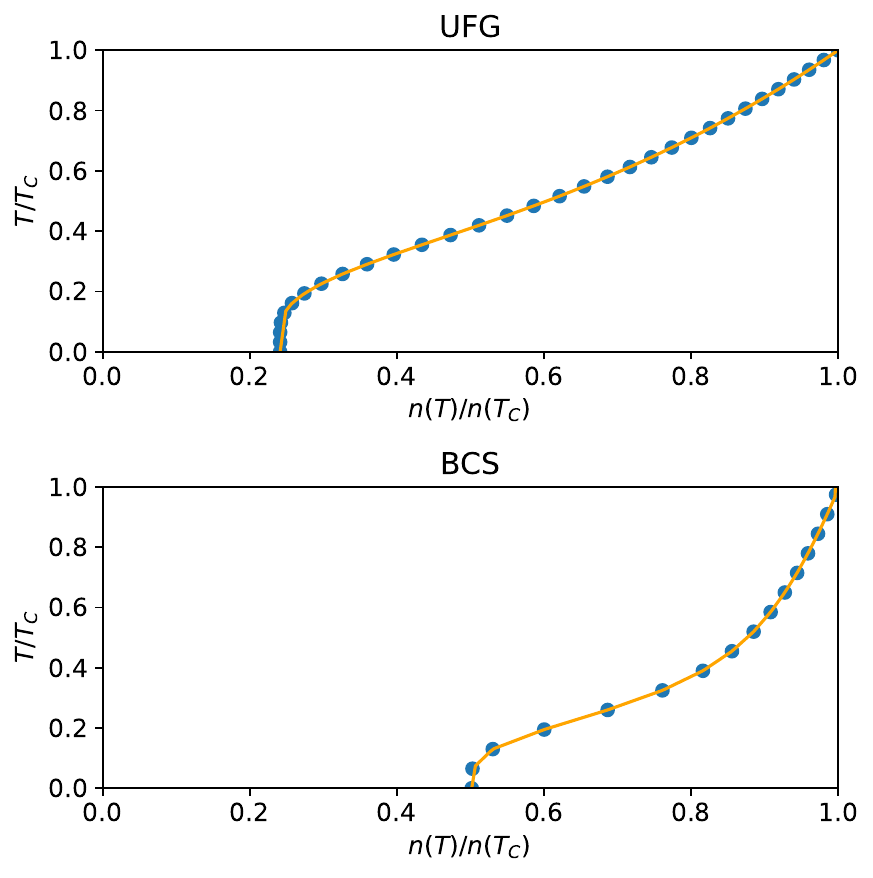}
  \caption[Thermometer calibration curve]{%
  The relation between the vortex core density and the gas temperature for \gls{UFG} (top) and \gls{BCS} (bottom) regimes. These calibration curves were used to assign the local effective temperatures, as presented in Fig.~8.
  }
	\label{fig:calibration}
\end{figure}

In Fig.~\ref{fig:calibration}, we provide the relation between vortex core density (normalized to the bulk density) and gas temperature. These curves were used for assign the local temperature of vortex cores, as shown in movies \verb|Supplementary movie 5| and \verb|Supplementary movie 6|.

\end{document}